\documentclass[acmsmall]{acmart}
\acmConference[]{}{}{}
\acmYear{}
\acmISBN{} 
\acmDOI{} 
\startPage{1}

\setcopyright{none}

\usepackage{amsmath}
\usepackage{caption}
\usepackage{subcaption}
\usepackage{listings}

\lstdefinelanguage{JavaScript}{
  morekeywords=[1]{break, continue, delete, else, for, function, if, in,
    new, return, this, typeof, var, void, while, with},
  morekeywords=[2]{false, null, true, boolean, number, undefined,
    Array, Boolean, Date, Math, Number, String, Object},
  morekeywords=[3]{eval, parseInt, parseFloat, escape, unescape},
  sensitive,
  morecomment=[s]{/*}{*/},
  morecomment=[l]//,
  morecomment=[s]{/**}{*/}, 
  morestring=[b]',
  morestring=[b]"
}[keywords, comments, strings]

\newcommand{\prg}[1]{\textbf{#1}}
\newcommand{\ignore}[1]{}

\title{Reliable Actors with Retry Orchestration}

\author{Olivier Tardieu}
\affiliation{
  \institution{IBM Research}
  \streetaddress{1101 Kitchawan Road}
  \city{Yorktown Heights}
  \state{New York}
  \postcode{10598}
  \country{United States of America}
}
\email{tardieu@us.ibm.com}

\author{David Grove}
\orcid{0000-0002-3265-7125}
\affiliation{
  \institution{IBM Research}
  \streetaddress{1101 Kitchawan Road}
  \city{Yorktown Heights}
  \state{New York}
  \postcode{10598}
  \country{United States of America}
}
\email{groved@us.ibm.com}

\author{Gheorghe-Teodor Bercea}
\affiliation{
  \institution{IBM Research}
  \streetaddress{1101 Kitchawan Road}
  \city{Yorktown Heights}
  \state{New York}
  \postcode{10598}
  \country{United States of America}
}
\email{doru.bercea@amd.com}

\author{Paul Castro}
\affiliation{
  \institution{IBM Research}
  \streetaddress{75 Binney Street}
  \city{Cambridge}
  \state{Massachusetts}
  \postcode{02142}
  \country{United States of America}
}
\email{castrop@us.ibm.com}

\author{Jaroslaw Cwiklik}
\affiliation{
  \institution{IBM Research}
  \streetaddress{3039 E Cornwallis Rd}
  \city{Research Triangle Park}
  \state{North Carolina}
  \postcode{27709}
  \country{United States of America}
}
\email{cwiklik@us.ibm.com}

\author{Edward Epstein}
\affiliation{
  \institution{IBM Research}
  \streetaddress{1101 Kitchawan Road}
  \city{Yorktown Heights}
  \state{New York}
  \postcode{10598}
  \country{United States of America}
}
\email{eae@us.ibm.com}

\ccsdesc[500]{Software and its engineering~Error handling and recovery}
\keywords{distributed systems, actors, fault tolerance}

\begin{abstract}
Cloud developers have to build applications that are resilient to failures and
interruptions. We advocate for a fault-tolerant programming model for the cloud
based on actors, retry orchestration, and tail calls. This model builds upon
persistent data stores and messages queues readily available on the cloud. Retry
orchestration not only guarantees that (1) failed actor invocations will be
retried but also that (2) completed invocations are never repeated and (3) it
preserves a strict happen-before relationship across failures within call
stacks. Tail calls can break complex tasks into simple steps to minimize
re-execution during recovery. We review key application patterns and failure
scenarios. We formalize a process calculus to precisely capture the mechanisms
of fault tolerance in this model. We briefly describe our implementation. Using
an application inspired by a typical enterprise scenario, we validate
the functional correctness of our implementation and assess the impact
of fault preparedness and recovery on performance.

\end{abstract}

\begin{document}

\maketitle

\section{Introduction}
\label{sec:intro}
Clouds are complex distributed systems with many components each with their own
points of failure~\cite{10.1145/1807128.1807161,10.1016/j.jnca.2016.08.010}.
Cloud developers have to build applications that are resilient to failures and
interruptions or face the risk of downtime and data loss. Many cloud
applications leverage existing cloud services. While some services like
persistent data stores can help achieving fault tolerance, others offer no or
weak fault tolerance guarantees. Naively composing such services can be
disastrous.

Distributed programming models often strive to achieve or get close to
transparent fault-tolerance. Because of hardware failures or software errors,
tasks can fail and data can be lost. To tolerate faults, distributed runtime
systems have to retry failed tasks and rebuild lost data, typically from a
checkpoint or a log. Transparent fault-tolerance therefore implies that making
multiple attempts at running a task should not be observably different from
running the task exactly once. As a result, many programming models require that
tasks (or just unfinished tasks) have no observable side effects, or that
retries of tasks do not repeat prior side effects. They may help by buffering
side effects then atomically applying all side effects at task completion time.
They may also attempt to deduplicate or roll back the side effects produced by
retries. All of these approaches require tasks to have certain properties---task
may be pure, stateless, deterministic, idempotent, reversible, etc.---and side
effects to be managed by the runtime system.

In this work, we target applications built by composing existing cloud services
and new code. We can decide a programming model for the new code and build a
runtime system to bridge new and existing application components, but we cannot
constrain existing components and their invocation. In general, we cannot assume
or control much about the side effects of service invocations. Moreover, we
observe that fault-tolerance, transparent or not, does not mean that faults
should not affect the behavior of an application. For example, many
applications make time-sensitive decision. If a decision is delayed for too long
because of a fault, the outcome must be different. In short, we also cannot
assume that the side effects of tasks will be the same with each execution
attempt.

Because of this harsh reality, we believe a novel programming model for
fault-tolerant distributed application is needed. This programming model cannot
achieve transparent fault-tolerance, as it cannot magically mask or undo side
effects from failed tasks, and cannot coalesce side effects within or across
retries. The goal of this programming model therefore must be to facilitate
authoring fault recovery code.
In this work we introduce, formalize, and evaluate KAR: a novel programming
model, methodology, and runtime system for building fault-tolerant cloud
applications. KAR builds upon the actor programming model. Applications consist
of invocations of actor methods, which we refer to as tasks for brevity.
KAR makes the following contributions:
\begin{itemize}
\item KAR makes it easy to split complex tasks into sequences of simple tasks
using tail calls, so as to minimize re-execution upon failure.
\item KAR automates and orchestrates retries ensuring that not only are failed
tasks retried, but also that successful tasks are never repeated.
\item A strict happen-before relationship is enforced between attempts at
executing a task \emph{transitively including all subtasks of this task
(blocking nested invocations)}, so that the ``past'' never leaks into the
``present''.
\item KAR automatically persists pending task parameters and results
into message queues but does not prescribe how to persist actor state
or interface with external services.
\end{itemize}
Tail calls make the transition from one task to the next atomic, ensuring a
failure never results in retrying both caller and callee. Moreover, consecutive
invocations of the same actor retain the actor lock. KAR preserves the lock
across failures ensuring that the interrupted chain is resumed first.

Retry orchestration frees developers from implementing retries and reasoning
about redundant or overlapping retries. At the developer's discretion,
non-failed subtasks of a failed task may be canceled if still pending, preempted
if already running, or awaited if neither canceled nor preempted.

KAR does not assume that all application state is funneled into a common
persistent store, with the luxury of transactions or atomic updates. On
the contrary, KAR obeys the separation principle of microservice architectures.
letting actors directly interface with data stores and stateful services.

However, KAR does assume that an application component can be forcefully disconnected from
the services it interacts with, eventually terminating all on-going or pending
invocations involving it. This requirement can be satisfied in
different ways: forcefully closing connections, killing processes, flushing
queues, implementing epochs, or waiting for quiescence or timeouts.
Importantly, this requirement does not necessitate coordination across multiple
services. It makes it possible to generalize the happen-before guarantee
from tasks to service invocations.

Forceful disconnection combined with tail calls and retry orchestration make it
possible to productively compose actors and cloud services (fault-tolerant or
not) into fault-tolerant applications as illustrated in Section~\ref{sec:model}
with a small example of an actor wrapping a key-value store and in
Section~\ref{sec:app} via a workflow drawn from an enterprise application.

Like Resilient X10~\cite{Res-X10:TOPLAS,Crafa2014:ECOOP}, KAR automatically
preserves ordering dependencies across failures but unlike X10 it automatically
retries failed tasks. Like Durable Functions~\cite{Burckhardt:OOPSLA21}, given
deterministic tasks KAR can produce deterministic outcomes irrespective of
failures. But unlike Durable Functions, KAR does not require orchestration
functions to be deterministic. Like Reliable State
Machines~\cite{RSM:ecoop2019}, KAR relies on reliable message queues to connect
application components and build a replay-able log, but unlike Reliable State
Machines, KAR permits applications to persist state outside of this log. Popular
distributed actor runtimes like Akka~\cite{akka} and Ray~\cite{222605} redeliver
an invocation request to a failed actor as soon the actor has recovered. They
permit the retried invocations to run concurrently with subtasks of the original
invocation. At best, this overlap complicates analyzing logs, at worst it can
cause subtle races and deadlocks. KAR eliminates this overlap.

The main contributions of this paper are:
\begin{itemize}
\item We introduce KAR (Section~\ref{sec:model}) and formalize the core
semantics of KAR to precisely formulate and establish its fault tolerance
guarantees (Section~\ref{sec:semantics}).
\item We succinctly describe an implementation of KAR (Section~\ref{sec:impl})
and an enterprise cloud application built using KAR (Section~\ref{sec:app}).
\item We empirically validate the functional correctness of our
implementation and assess the impact of fault preparedness and
recovery on performance (Section~\ref{sec:eval})
\end{itemize}

\section{Programming Model}
\label{sec:model}
The KAR programming model is not tied to a specific programming language. In
this section, we demonstrate KAR using JavaScript as the base language. So far
we have developed KAR applications using JavaScript, TypeScript, Python, and
Java.

KAR applications are made of actors. A \prg{Latch} actor type may be implemented
by a JavaScript class:
\begin{lstlisting}[columns=fullflexible]
class Latch {
  async activate() { this.v = 0 }
  async set(v) { this.v = v }
  async get() { return this.v }
}
\end{lstlisting}
The state of an actor instance, \prg{this.v} in this example, is meant to be
private and only accessed through instance methods. The \prg{activate} method
essentially functions as a constructor. It is implicitly invoked by KAR
at actor instantiation time. Because many programming languages have
strong opinions about what constructors can and cannot do, we prefer this
portable idiom to constructors.

An actor is identified by its type and a unique instance id. The
\prg{actor.proxy} method synthesizes a reference to an actor instance:
\begin{lstlisting}[columns=fullflexible]
let ref = actor.proxy('Latch', 'myInstance')
\end{lstlisting}
Multiple invocations of \prg{actor.proxy} with the same parameters synthesize
equivalent references, i.e., references to the same actor instance. The
\prg{actor.proxy} method does not instantiate actors. Actors are instantiated
implicitly when first invoked.

Actors methods are always invoked indirectly:
\begin{lstlisting}[columns=fullflexible]
await actor.call(actor.proxy('Latch', 'myInstance'), 'set', 42)
\end{lstlisting}
This indirection makes it possible for the KAR runtime to persist and possibly
retry invocation requests. It also abstracts away the application topology: the
caller and callee may be running in different application components.

Because IOs in JavaScript are normally asynchronous and because actor methods
are meant to be remotely invocable, we uniformly make every actor method
\prg{async} and \prg{await} every call. The \prg{async} and \prg{await} keywords
are JavaScript idiosyncrasies. The reader unfamiliar with these keywords is
encouraged to just ignore them. The equivalent Java code would not have them.

Invocation using \prg{actor.call} are blocking. The call returns the result from
the callee. Non-blocking actor invocations are written:
\begin{lstlisting}[columns=fullflexible]
await actor.tell(actor.proxy('Latch', 'myInstance'), 'set', 42)
\end{lstlisting}
This call only waits for the invocation request to be acknowledged by the KAR
runtime but does not wait for a result.

Exceptions in \prg{actor.call} are propagated from callees to callers where they
may be caught. Exceptions in \prg{actor.tell} are logged and discarded. KAR also
supports \emph{reminders}\/, i.e., time-delayed and possibly periodic variants
of \prg{actor.tell}.

\subsection{Failures and Persistence}

Actors can fail, at once losing the in-memory state of the instance and
interrupting all the method invocations in progress. However, the parameters of
queued method invocations including invocations in progress at the time of a
failure are not lost, being automatically persisted by KAR. A failed actor is
automatically recreated by KAR if invoked again whether because of a retried
invocation or a novel invocation.

KAR offers a persistence API for actors to save their state. 
\begin{lstlisting}[columns=fullflexible]
class PersistentLatch {
  async activate() { this.v = await actor.state.get(this, 'v') | 0 }
  async set(v) { this.v = v; await actor.state.set(this, 'v', this.v) }
  async get() { return this.v }
}
\end{lstlisting}
Because \prg{activate} is called both on construction and reconstruction upon
failure, it should restore the in-memory state of the instance from its
persisted state if any, or if none initialize the instance state, loading
\prg{0} into \prg{this.v} in this example.

Actors however are free to choose if, when, and how to persist their state.
KAR's retry orchestration guarantees are not predicated on the use of the
builtin persistence API.

\subsection{Nested Calls, Reentrancy, Retry Orchestration}

Nested calls, i.e., blocking actor calls originating from an actor, must pass an
extra \prg{this} argument as the first parameter of the call, shifting other
\prg{actor.call} arguments to the right. The identity of the caller is required
by the runtime to permit reentrancy and to correctly orchestrate retries.
\begin{lstlisting}[columns=fullflexible]
class A {
  async main(v) { await actor.call(this, actor.proxy('B', 'b'), 'task', v, this) }
  async callback(v) { console.log(v) }
}
class B {
  async task(v, a) { await actor.call(this, a, 'callback', v) }
}
await actor.call(actor.proxy('A', 'a'), 'main', 42)
\end{lstlisting}
KAR actors are single-threaded and reentrant. Invocations of an actor are queued
and processed one at a time in queue order, except for nested invocations. If a
method of an actor calls a method of the same actor using only a stack of nested
calls (\prg{actor.call} with extra \prg{this} argument), then the nested
invocation runs immediately bypassing the queue. In this example code, the call
to \prg{callback} does not deadlock but runs while \prg{main} is suspended
waiting for \prg{task} to return.

\begin{figure}
\includegraphics[trim=280 220 0 10,clip,width=.9\columnwidth]{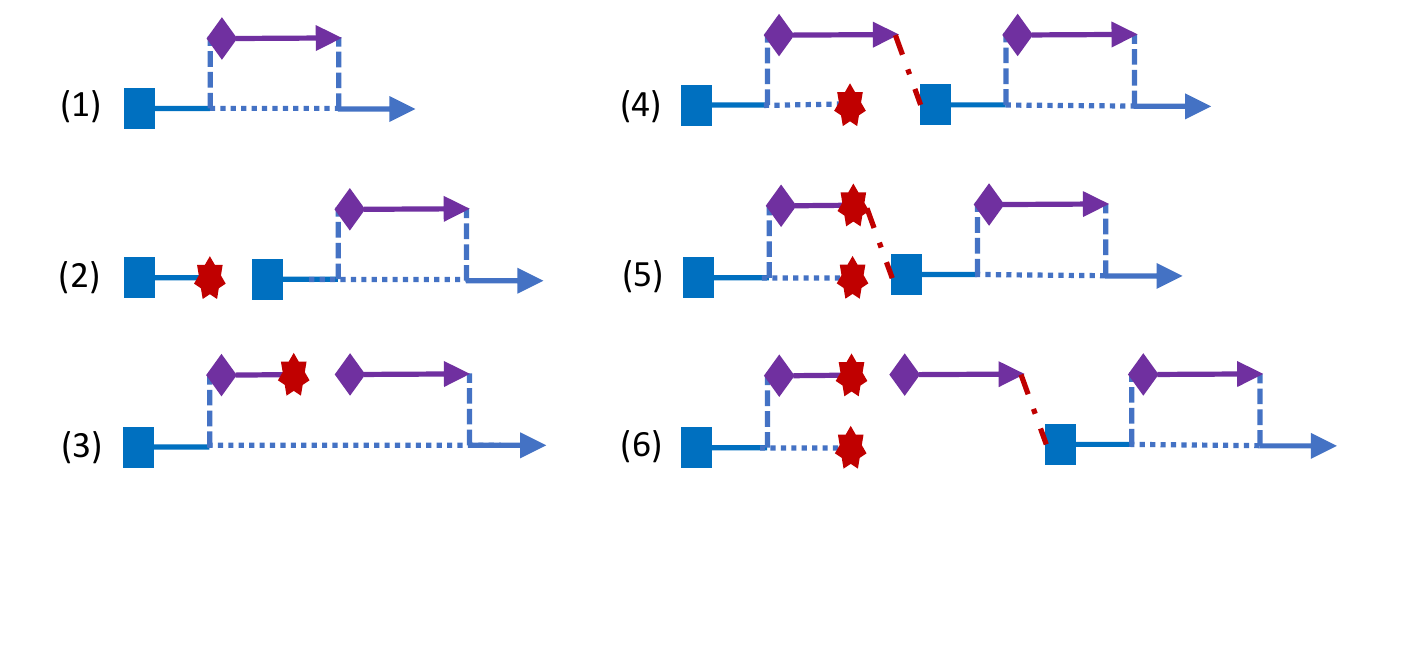}
\caption{Retry orchestration for a nested call.}
\vspace{-0.15in}
\label{fig:timelines}
\Description{Time line diagrams showing possible execution scenarios for nested
calls with and without failures.}
\end{figure}

Figure~\ref{fig:timelines} shows the possible execution timelines for a nested
call without failures (1) or with a single failure (2-7). A first actor method
represented by a square calls a second method represented by a diamond. These
could be the \prg{main} and \prg{task} methods of the previous example. A line
ending with an arrow depicts a complete execution of the task, whereas a star
depicts a failure interrupting the task, and a stop sign an intentional
preemption by the runtime.

A failure may hit the caller before the call, resulting in a retry of the caller
(2). It may hit the callee only, resulting in a retry of the callee (3). A
failure may also hit the caller while it is waiting for the callee (4-7).
Scenarios (4-5) describe the possible recovery strategies for a failure only
affecting the caller. In (4), the callee runs to completion, whereas in (5) the
callee is preempted. Scenarios (6-7) describe the possible strategies for a
joint failure. In (6), the caller is retried, eventually reinvoking the callee.
In (7), the failed callee is retried first, then the caller as in (6). A failure
may also hit the caller after placing the call but before the callee has started
leading to a similar choice (omitted from the figure for brevity).

KAR does not prescribe the choice of (4) vs. (5) or (6) vs. (7). We formalize
all these options. Preempting a running task or canceling a pending task may or
may not be feasible or desirable. In (4-7) as illustrated by the oblique dashed
red line, KAR delays the retry of the caller waiting either for the callee to
complete (4), to be preempted (5), to be known to have failed (6), or to be
retried (7). In the absence of failures, a call stack is a single logical thread
of execution. KAR ensures that the decision on the callee's fate and possible
execution always \emph{happen before} the retry of the caller to preserve this
logic upon failure.

\begin{figure}
\includegraphics[clip,width=.8\columnwidth]{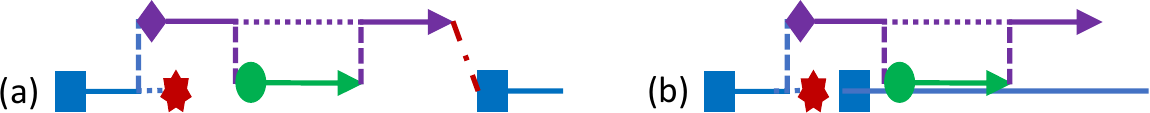}
\caption{Reentrancy (a) with and (b) without happen-before.}
\label{fig:reentrancy}
\Description{Time line diagrams showing possible reexecutions of a reentrant
call with and without the enforcement of happen-before.}
\vspace{-0.15in}
\end{figure}

This guarantee is critical for safe reentrancy. Consider what would happen if
\prg{main} fails after invoking \prg{task}. Figure~\ref{fig:reentrancy}(a)
depicts the executions allowed by KAR where happens before ensures that
\prg{main} will only start executing after \prg{task} has completed.
Figure~\ref{fig:reentrancy}(b) shows what could happen if \prg{main} is retried
without waiting, with \prg{main} and \prg{callback} now executing concurrently.
Queuing \prg{callback} to run later does not help as that would result in a
deadlock when \prg{task} is reinvoked by the retried \prg{main}.

\subsection{Tail Calls}

Nested calls require precise handling from the runtime and careful thinking from
the developer to account for the possible failure scenarios (3-7) when both
invocations have been issued already, but neither has completed yet. However, if
the invocation of the callee is the last thing the caller wants to do, we can
replace the nested call (\prg{actor.call}) with a tail call
(\prg{actor.tailCall}). In essence, a tail call simultaneously completes the
caller's invocation while issuing the invocation of the callee.

For instance, tail calls make it possible to reliably increment a counter in a
key-value store with only \prg{get} and \prg{set} methods:
\begin{lstlisting}[columns=fullflexible]
class Accumulator {
  async get() { return await store.get('key') }
  async set(value) { await store.set('key', value); return 'OK' }
  async incr() {
    let value = await store.get('key')
    return await actor.tailCall(this, 'set', value+1)
  }
}
\end{lstlisting}
The \prg{incr} method reads the value from the store then makes a tail call to
the \prg{set} method to store the incremented value. Because all tail calls
originate from actor instances, the first argument to \prg{actor.tailCall} is
always \prg{this}.

A chain of tail calls returns the return value of the last call in the chain. In
this example, a caller making a blocking \prg{actor.call} to \prg{incr} remains
blocked when \prg{incr} returns a tail call expression. This expression is
recognized by the KAR runtime, which atomically records the completion of
\prg{incr} and the request to invoke \prg{set}. The caller of \prg{incr} is
eventually unblocked when it receives the \prg{OK} value returned upon
completion of the \prg{set} method.

A tail call to the same actor retains the actor lock, whereas a tail call to a
different actor releases the lock. In this example, the lock is retained between
the execution of \prg{incr} and the matching execution of \prg{set}. As a
result, no \prg{store.get} or \prg{store.set} operation from a different call to
\prg{get}, \prg{set}, or \prg{incr} can be interleaved between the
\prg{store.get} and \prg{store.set} operations belonging to the \prg{incr} call.
This guarantees holds even upon failure as KAR persists the lock. In short,
concurrent invocations of \prg{incr} from different callers are serialized.

Thanks to the tail call, a failure may either interrupt \prg{incr} or \prg{set}
but not both. If \prg{incr} is interrupted, the \prg{store.get} operation may be
repeated. Because \prg{set}, hence \prg{store.set}, has not run yet, the read
value will remain the same. If on the other hand, \prg{set} is interrupted, the
\prg{store.set} operation may be repeated, but because the read value has been
cached as an invocation parameter to \prg{set}, the same value will be written
every time by \prg{store.set}. In all cases, the value will be incremented
exactly once by the time the caller of \prg{incr} receives the \prg{OK} return
value.

This is not the full story though. KAR ensures executions of \prg{set} cannot
overlap, \prg{store.set} invocations however are not KAR tasks. In particular,
there is no guarantee a priori that a failure of \prg{set} interrupts
\prg{store.set}. In principle, a \prg{store.set} invocation could linger,
possibly delayed by the network, and execute well out of order, including after
another successful \prg{incr} call. In other words, even if the \prg{get} and
\prg{set} tasks from concurrent \prg{incr} tasks cannot be interleaved by KAR,
delayed store invocations could still corrupt the store, which is why KAR
requires and relies on forceful disconnection. By forcing the \prg{store.set}
invocation from a failed \prg{set} task to settle (complete or abort) before
retrying the \prg{set} task, KAR can extend its ordering guarantees from tasks
to store operations, guaranteeing correct increments.

While it may be tempting to read and write from a single method body or tempting
to replace the tail call with a nested call, neither works. Consider these two
incorrect variants:
\begin{lstlisting}[columns=fullflexible]
async incr() { await store.set('key', await store.get('key')+1); return 'OK' }
async incr() { return await actor.call(this, 'set', await store.get('key')+1) }
\end{lstlisting}
The top method may fail after \prg{store.set} but before return, resulting in
multiple increments upon retry. The bottom method, which uses a nested call
rather than a tail call, may fail after the \prg{set} call before returning,
also leading to multiple increments upon retry. Unsurprisingly, there
is no reliable way to implement a fault-tolerant increment without an atomic
operation such as \prg{actor.tailCall}.

\subsection{Benefits}
\label{sec:benefits}

KAR augments a traditional actor-based programming model in two fundamental
ways:
\begin{itemize}
\item KAR keeps track of call stacks, so it can 1) enable reentrant calls and 2)
orchestrate, i.e., delay, the retry of a caller until after the callee's fate
has been decided: either completed without failure, preempted, known to have
failed, or successfully retried.
\item KAR introduces tail calls. A tail call atomically completes a call while
issuing a next call. In addition, a tail call retains the lock on the caller if
the caller is also the callee.
\end{itemize}

Tail calls greatly simply reasoning about failures when applicable, while retry
orchestration makes reasoning about failures tractable when tail calls are not
applicable by ensuring past call stacks cannot overlap with retries of these
call stacks.

Tail calls have two important, somewhat distinct use cases:
\begin{itemize}
\item Tail calls permit breaking complex methods into series of simple steps,
simple methods chained via tail calls, enabling a divide-and-conquer approach to
fault tolerance. Regular calls may also be used to break complex code. However,
regular calls complicate reasoning about failures, as callers and callees may or
may not fall victim of the same failure. Regular calls also complicate
concurrency control even in the absence of failure, which is avoided by tail
calls thanks to their locking behavior (irrespective of failures).
\item 
Tail calls enforce a state-machine-like transition discipline not just within
one actor but across actors. An actor may tail call other actors. With KAR,
actors and state machines are orthogonal concepts. Actors may represent orders,
payments, and shipments. Chains of tail calls can implement business processes
like receiving an order and processing a payment.
\end{itemize}

Tail calls encode a key transactional pattern without bearing the cost of more
general transactional mechanisms, both in terms of performance and productivity.
A tail call is a single message that semantically is both a request and a
response, rather than a collection of messages that must be emitted at once.
There is no concept of aborted transaction or rollback for the developer to
understand. Critically, tail calls enable developers to express these patterns
in KAR even when the underlying cloud services or APIs being composed lack
transactional support or retry capabilities as illustrated with the \prg{incr}
method code.

\section{Formal Semantics}
\label{sec:semantics}
\makeatletter
\providecommand*{\xmapstofill@}{\arrowfill@{\mapstochar\relbar}\relbar\rightarrow}
\providecommand*{\xmapsto}[2][]{\ext@arrow 0395\xmapstofill@{#1}{\,#2\,}}
\makeatother

\newcommand{\rul}[2]{\ensuremath{\begin{array}{c}{#1}\\\hline{#2}\end{array}}}
\newcommand{\spc}{\hspace{6mm}}
\newcommand{\short}{\hspace{3mm}}
\newcommand{\alt}{\hspace{1mm}|\hspace{1mm}}
\newcommand{\doubleplus}{\mathbin{+\!\!+}}
\newcommand{\prd}[1]{\textbf{#1}}

We first formalize method invocations~(\ref{sec:sem-base}). We specify the
syntax of terms that encode the possible points in the execution of a method and
several forms of transitions that encode possible execution steps. We then
specify the semantics of KAR by mapping method invocations to logical processes
and using messages to transport invocation requests and responses among
them~(\ref{sec:sem-distributed}). We define failures~(\ref{sec:sem-failures}).
We specify the \prd{runnable} predicate that decides when pending invocation
requests may be (re-)executed~(\ref{sec:sem-runnable}). We formalize KAR's
guarantees (\ref{sec:sem-guarantees}), cancellation, and preemption
(\ref{sec:sem-cancel}).

\subsection{Base Language Specification}
\label{sec:sem-base}

We assume a fixed but arbitrary program and abstract most of its syntax and
semantics. We use the following alphabet:
\begin{center}
\begin{tabular}{rlrlrl}
actor reference: & $a$ &
method name: & $m$ &
value: & $v$ \\
actor state: & $p$ &
request id: & $i$ &
sequel: & $s$
\end{tabular}
\end{center}
A point in the execution of a method is a pair $T/p$ where $T$ denotes the code
that remains to be executed and $p$ denotes the state of the actor the method is
running on.
\begin{equation*}
T ::= m(v) \alt v \alt s \alt a.m(v) \triangleright s \alt v \triangleright s \alt a.m(v) \wr s \tag {term}
\end{equation*}
The term $m(v)$ denotes the initial method invocation including the method name
$m$ and parameter $v$. The term $v$ denotes the return value. The term $s$
denotes an intermediate point in the method execution, more precisely the code
remaining to be executed combined with the local state (local variables), in
short the ``sequel''. The term $a.m(v)\triangleright s$ denotes a nested method
invocation (\prg{actor.call}) where $a.m(v)$ denotes the callee and $s$ denotes
the remainder of the caller to execute once the nested invocation has completed.
The term $v\triangleright s$ denotes the reception of a result $v$ from a nested
invocation with $s$ denoting the remainder of the caller. The term $a.m(v)\wr s$
denotes an asynchronous method invocation (\prg{actor.tell}), where $s$ denotes
the remainder of the caller that may execute concurrently with the callee.

We assume the program is specified as a set of valid transitions with forms:
\begin{center}
\begin{minipage}{.5\textwidth}
\begin{align*}
m(v)/p &\rightarrow s/p \tag{begin} \\
s/p &\rightarrow v/p \tag{end} \\
s/p &\rightarrow s'/p' \tag{step} \\
v \triangleright s/p &\rightarrow s'/p \tag{return}
\end{align*}
\end{minipage}%
\begin{minipage}{.5\textwidth}
\begin{align*}
s/p &\rightarrow a.m(v) \triangleright s'/p \tag{call} \\
s/p &\rightarrow a.m(v) \wr s'/p \tag{tell} \\
s/p &\rightarrow a.m(v)/p \tag{tail-call} \\
\end{align*}
\end{minipage}
\end{center}
The execution of a method starts with a (begin) transition. Because our focus is
on retry and ordering guarantees, we can abstract all the typical constructs of
an imperative programming language such as sequences, conditionals, loops, or
local variables in a single (step) form. The remaining forms permit method
invocations (call, tell, tail-call), return a value (end), and receive the
result of a nested invocation (return).

For example a \prg{getset} method of a \prg{Latch} actor that updates the actor
state returning the previous value, may be specified as the following infinite
set of transitions (for all $v$, for all $p$) assuming values and actor states have
the same domain:
\[ \text{getset}(v)/p\rightarrow\text{in}_v/p \spc\spc
\text{in}_v/p\rightarrow\text{out}_p/v \spc\spc \text{out}_p/v\rightarrow p/v \] The
families of sequels $\text{in}_v$ and $\text{out}_p$ represent intermediate
points in the execution of the method capturing both its progress (before or
after the value swap) and the local state of the method (the input or output
value).

This set of transitions is not an execution semantics. It does not specify how
to chain execution steps. This is simply an abstraction of a source code
designed to be language-neutral and focused on the features---actors and method
invocations---that matter to KAR's semantics. Similarly we do not specify how
actor instances may be derived, e.g., from classes. Concretely, this set of
transitions may be generated from a higher-level specification including
control-flow constructs, a data model, and a mechanism to map actor references
to sets of methods, for example by breaking actor references into a tuple (class
type, instance id).

\subsection{Message-Passing Semantics}
\label{sec:sem-distributed}

Each method invocation runs in its own logical process. Processes communicate by
means of invocation request and response messages. Processes running method
invocations on the same actor reference share the actor state, i.e., the ability
to read and write this shared state.

First we introduce some terms. A request id $i$ is an identifier. A
return address $r$ is an optional request id.
\begin{center}
\begin{minipage}{.5\textwidth}
\begin{align*}
M &::= i\xmapsto{r} a.m(v) \alt i\xmapsto{r} v \tag{message} \\
F &::= (M,M',...) \tag{flow} \\
P &::= s \alt i \triangleright s \tag{process}
\end{align*}
\end{minipage}%
\begin{minipage}{.5\textwidth}
\begin{align*}
E &::= \{i \xmapsto{a} P,...\} \tag{ensemble} \\
S &::= \{a \mapsto p,...\} \tag{persistent state} \\
R &::= F,E,S \tag{runtime state}
\end{align*}
\end{minipage}
\end{center}
A message $M$ is a 3-tuple consisting of a request id $i$, a return address $r$,
and either a method invocation $a.m(v)$ or a return value $v$. The return
address is the request id for the caller for a nested invocation, blank for an
asynchronous invocation.

A flow $F$ is a possibly empty, ordered list of messages. List concatenation is
written $F\doubleplus F'$. To keep the syntax of our semantics simple, we
formalize communications between actors as a unique flow, i.e., messages are
totally ordered. The order of messages however is only tested in rule (leftmost)
in Section~\ref{sec:sem-runnable}, which identifies the oldest invocation
request for a given actor reference $a$. Consequently, the position of a
response message is irrelevant, as is the relative position of request messages
sent to distinct actors.

A persistent state $S$ is a map from actor references $a$ to the states $p$ of
these actors. We assume there is a default empty actor state, meaning for
instance that the empty map $\emptyset$ maps every actor reference to the empty
actor state.

A process $P$ is either a sequel $s$ or a guarded sequel $i \triangleright s$. A
guarded sequel denotes a process waiting for the result of a nested invocation
with id $i$.

An ensemble $E$ is a map from request ids $i$ to processes $P$ tagged with actor
references $a$. The tag denotes the actor this process is running on. The union
of maps with disjoint key sets is written $E\uplus E'$.

\begin{figure*}
\begin{equation*}
\tag{begin}
\rul{\prd{runnable}(i, F \doubleplus (i \xmapsto{r} a.m(v)) \doubleplus F') \spc m(v)/p \rightarrow s/p}
  {F\doubleplus(i \xmapsto{r} a.m(v))\doubleplus F',E,S\uplus\{a\mapsto p\} \Rightarrow F\doubleplus(i \xmapsto{r} a.m(v))\doubleplus F',E\uplus\{i \xmapsto{a} s\},S\uplus\{a\mapsto p\}}
\end{equation*}
\begin{equation*}
\tag{step}
\rul{s/p \rightarrow s'/p'}
  {F,E\uplus\{i \xmapsto{a} s\},S\uplus\{a\mapsto p\} \Rightarrow F,E\uplus\{i \xmapsto{a} s'\},S\uplus\{a\mapsto p'\}}
\end{equation*}
\begin{equation*}
\tag{end}
\rul{s/p \rightarrow v/p}
  {F\doubleplus(i \xmapsto{r} a.m(v))\doubleplus F',E\uplus\{i \xmapsto{a} s\},S\uplus\{a\mapsto p\} \Rightarrow F\doubleplus F'\doubleplus(i \xmapsto{r} v),E,S\uplus\{a\mapsto p\}}
\end{equation*}
\begin{equation*}
\tag{call}
\rul{i'\text{ fresh} \spc s/p \rightarrow a'.m(v) \triangleright s'/p}
  {F,E\uplus\{i \xmapsto{a} s\},S\uplus\{a\mapsto p\} \Rightarrow F\doubleplus(i' \xmapsto{i} a'.m(v)),E\uplus\{i \xmapsto{a} i' \triangleright s'\},S\uplus\{a\mapsto p\}}
\end{equation*}
\begin{equation*}
\tag{tell}
\rul{i'\text{ fresh} \spc s/p \rightarrow a'.m(v) \wr s'/p}
  {F,E\uplus\{i \xmapsto{a} s\},S\uplus\{a\mapsto p\} \Rightarrow F\doubleplus(i' \mapsto a'.m(v)),E\uplus\{i \xmapsto{a} s'\},S\uplus\{a\mapsto p\}}
\end{equation*}
\begin{equation*}
\tag{return}
\rul{v \triangleright s/p \rightarrow s'/p}
  {F\doubleplus(i' \xmapsto{i} v)\doubleplus F',E\uplus\{i \xmapsto{a} i' \triangleright s\},S\uplus\{a\mapsto p\} \Rightarrow F\doubleplus  F',E\uplus\{i \xmapsto{a} s'\},S\uplus\{a\mapsto p\}}
\end{equation*}
\begin{equation*}
\tag{tail-other}
\rul{a \neq a' \spc s/p \rightarrow a'.m(v)/p}
  {F\doubleplus(i \xmapsto{r} a.m'(v'))\doubleplus F',E\uplus\{i \xmapsto{a} s\},S\uplus\{a\mapsto p\} \Rightarrow F\doubleplus F'\doubleplus(i \xmapsto{r} a'.m(v)),E,S\uplus\{a\mapsto p\}}
\end{equation*}
\begin{equation*}
\tag{tail-self}
\rul{s/p \rightarrow a.m(v)/p}
  {F\doubleplus(i \xmapsto{r} a.m'(v'))\doubleplus F',E\uplus\{i \xmapsto{a} s\},S\uplus\{a\mapsto p\} \Rightarrow F\doubleplus(i \xmapsto{r} a.m(v))\doubleplus F',E,S\uplus\{a\mapsto p\}}
\end{equation*}
\caption{Message-passing semantics.}
\Description{Message-passing semantics.}
\label{fig:semantics}
\end{figure*}

A runtime state $R$ is a 3-tuple consisting of a flow $F$, an ensemble $E$, and a persistent
state~$S$. We specify KAR's semantics as a transition system of the form $F,E,S
\Rightarrow F',E',S'$. The initial runtime state is made of a single request
message with the main invocation and no return address: $\{i \mapsto
m(v)\},\emptyset,\emptyset$. The rules of this semantics shown in
Figure~\ref{fig:semantics} are derived from the semantic forms of the base
program specification.

Rule (begin) starts the execution of a pending request if it is runnable and not
running already (because of the disjoint union of ensembles). The \prd{runnable}
predicate defined in \ref{sec:sem-runnable} encapsulates complex logic including
concurrency control and ordering constraints. Importantly for fault tolerance,
the request message remains in the flow at this time.

Rule (step) ensures that a running actor may only update its own state. No actor
can access the state of other actors.

Rule (end) atomically (1) discards the process at the end of the
invocation, (2) discards the request message, and (3) enqueues the response message with the return
value.

Rule (call) and (tell) allocate a fresh request id $i'$ (an id never used
before), attach this id to the nested method invocation request, and appends the
request to the tail of the flow. Rule (call) suspends the
caller with a guarded sequence and sets the message's return address,
whereas rule (tell) neither suspends the caller nor sets a return address.

Rule (return) resumes a guarded sequence by extracting the response message from
the flow.

For tail calls we distinguish calls to the same actor from calls to other
actors. In contrast with the (call) and (tell) rules, the tail call rules do not
introduce a fresh request id but reuse the id of the caller. Rule (tail-other)
atomically (1) removes the running process from the ensemble, (2) removes the
completed request from the flow, and (3) appends the tail call request to the
tail of the flow. Rule (tail-self) is almost the same except it inserts the tail
call message in the flow at the position of the removed message. This position
makes the tail call retain the logical lock on the actor irrespective of
failures and without affecting other requests.

\subsection{Failure Semantics}
\label{sec:sem-failures}

We specify failures by means of a single rule where $E\backslash A$ denotes the
ensemble $E$ with all entries labelled with $a\in A$ removed:
\begin{equation*}
\label{failure}
\tag{failure}
F,E,S \Rightarrow F,E\backslash A,S
\end{equation*}
Failures have no preconditions. They can happen at any time. A failure results
in the loss of all the method invocations running on a set of actors. Messages
and persistent state are not impacted. This rule reflects the nature of the
cloud platform KAR is targeting. Individual OS processes, containers, pods, or
nodes can fail. On the other hand, messages queues and data stores are
sophisticated systems orchestrating multiple distributed processes to provide a
level of redundancy. These systems can transparently tolerate and mitigate
failures up to a point, i.e., up to \emph{catastrophic} failures. The
catastrophic failure threshold depends on the particulars of given system
implementation and configuration. The KAR runtime is meant to provide
fault-tolerance guarantees in the absence of a catastrophic failure of the
message queue or data store.

While we do not model transient actor state, our semantics can be easily
extended to do so by dividing actor states into two components, one of which is
lost in the (failure) rule.

This fail-stop failure model does not fully capture the complexities of a
distributed runtime system where failures may be partial or transient and failure
detection is imperfect. We explain in Section~\ref{sec:impl} how we address
these issues by masking short transient failures and forcefully disconnecting
application components deemed to have failed.

\subsection{Runnable Invocations}
\label{sec:sem-runnable}

The semantic rules we introduced so far specifies how to execute method
invocations but not which invocations
to run. This is the purpose of the \prd{runnable} predicate. In a non-reentrant,
non-resilient, in-order actor system, an invocation is runnable if and only if
it is the oldest invocation enqueued on this actor.

To handle reentrancy, we first introduce the \prd{reachable} predicate defined by
induction. The invocation id of the oldest, i.e., leftmost, invocation of actor
$a$ is reachable from $a$ as well as any invocation of any actor $a'$ transitively nested in the leftmost invocation of $a$:
\begin{equation*}
\tag{leftmost}
\rul{i'\xmapsto{r'} a.m'(v') \not\in F}
  {\prd{reachable}(i, a, F \doubleplus (i \xmapsto{r} a.m(v)) \doubleplus F')}
\end{equation*}
\begin{equation*}
\tag{nested}
\rul{\prd{reachable}(i', a, F) \spc i \xmapsto{i'} a'.m(v) \in F}
  {\prd{reachable}(i, a, F)}
\end{equation*}

An invocation with id $i$ of actor $a$ is runnable if it is reachable from $a$, i.e., if
it is the oldest invocation of $a$ or nested in this invocation, and there is no
nested invocation with return address $i$ queued in the flow:
\begin{equation*}
\tag{runnable}
\rul{i \xmapsto{r} a.m(v) \in F \short \prd{reachable}(i, a, F) \short i'\xmapsto{i} a'.m'(v') \not\in F}
  {\prd{runnable}(i, F)}
\end{equation*}
In a failure-free execution, the latter condition is redundant. The only way for
a nested invocation to exist is for the caller to be running already, waiting
for the result from the callee. Because of failures however, the process for the
caller may be lost before the process for the callee completes. We say the
caller has to wait for the callee, in the sense that any retry of the caller $i$
has to wait for every callee $i'$ from any prior execution of the caller to
finish first. This condition embodies the happen-before edges illustrated in
Section~\ref{sec:model}.

\subsection{Formal Guarantees}
\label{sec:sem-guarantees}

Let $\Rrightarrow$ be the transitive, reflexive closure of $\Rightarrow$. KAR's
main retry guarantee may be formalized as follows:

\begin{theorem}
If $\{i\mapsto a.m(v)\},\emptyset,\emptyset\Rrightarrow F,E,S\Rrightarrow
F',E',S'$ and $E$ contains a process with id $i'$ and tag $a'$ and $F'$ contains
a request with id $i'$ then $\prd{reachable}(i, a', F')$.
\end{theorem}

Once a request starts running it remains reachable until it returns a result.
Importantly, there is no guarantee the request will ever be runnable again
because one attempt at running this request may make a nested call that never
returns, which would delay a retry indefinitely. This is arguably the desired
behavior as, in the absence of a failure, the invocation would have also been
stuck. Assuming the nested call eventually completes, the request becomes
runnable again, i.e., retry-able. In short, the happen-before guarantee has
priority over the retry guarantee.

The no retry after success guarantee (3.2),
no concurrent retries of a call or two calls in a chain of tail calls guarantee (3.3),
and happen-before guarantee (3.4) are safety properties:

\begin{theorem}
If $\{i\mapsto a.m(v)\},\emptyset,\emptyset\Rrightarrow F,E,S\Rrightarrow F',E',S'$ and
$F$ contains a response message with id $i'$ then $E'$ does not contain a process
with id $i'$.
\end{theorem}


\begin{theorem}
If $\{i\mapsto a.m(v)\},\emptyset,\emptyset\Rrightarrow F,E,S$ then $E$ contains at most one process with id $i'$.
\end{theorem}


\begin{theorem}
If $\{i\mapsto a.m(v)\},\emptyset,\emptyset\Rrightarrow F,E,S$ and
$F$ contains a request message with id $i'$ and return address $i''$ then $\prd{runnable}(i'', F)$ is false.
\end{theorem}

\begin{proof}
3.3 and 3.4 respectively follow from the definition of the ensemble of processes
and the \prd{runnable} predicate. 3.1 and 3.2 are established by induction over
the structure of the semantics. The key observations are (1) that the
\prd{reachable} predicate only depends on the prefix of the flow up to the
request under consideration so requests added to the tail of flow do not affect
earlier requests and (2) that execution in a call stack completes from the
inside out so there is no way for a request to disappear from the middle of a
call stack.
\end{proof}


\subsection{Optional Cancellation and Preemption}
\label{sec:sem-cancel}

\begin{figure*}
\begin{equation*}
\tag{cancel}
\rul{\prd{runnable}(i, F \doubleplus (i \xmapsto{i'} a.m(v)) \doubleplus F') \spc i'\xmapsto{a'}i\triangleright s\not\in E \spc i\xmapsto{a}T\not\in E}
  {F\doubleplus(i \xmapsto{i'} a.m(v))\doubleplus F',E,S \Rightarrow F\doubleplus F',E,S}
\end{equation*}
\begin{equation*}
\tag{preempt}
\rul{\prd{runnable}(i, F \doubleplus (i \xmapsto{i'} a.m(v)) \doubleplus F') \spc \prd{preemptable}(i, F \doubleplus (i \xmapsto{i'} a.m(v)) \doubleplus F') }
  {F\doubleplus(i \xmapsto{i'} a.m(v))\doubleplus F',E,S \Rightarrow F\doubleplus F',E\backslash i,S}
\end{equation*}
\caption{Cancellation and preemption.}
\Description{Cancellation and preemption.}
\label{fig:cancellation}
\end{figure*}

As discussed in Section~\ref{sec:model}, if desired, KAR may automatically
attempt to cancel pending nested invocations or preempt running nested
invocations to avoid unnecessary computations, i.e., the computation of a result
that is not needed anymore. Consequently, cancellation and preemption are only
applicable to nested invocation (\prg{actor.call}).

Because in a distributed system failure detection is not instantaneous, we
formalize cancellation and preemption as racing with normal execution. For
example, the preemption of a process may happen at any point in its execution or
not at all, once the conditions for preemption have been fulfilled.

Formally, we add either one of the two rules of Figure~\ref{fig:cancellation} to
the rules of Figure~\ref{fig:semantics}. Rule (cancel) removes a runnable nested
invocation request from the flow if there no process waiting for its result and
this invocation is not already running. Rule (preempt) generalizes (cancel)
using an auxiliary predicate:
\begin{equation*}
\tag{preemptable-root}
\rul{i'\xmapsto{a'} i\triangleright s \not\in E}
  {\prd{preemptable}(i, F \doubleplus (i \xmapsto{i'} a.m(v)) \doubleplus F', E)}
\end{equation*}
\begin{equation*}
\tag{preemptable-nested}
\rul{\prd{preemptable}(i, F \doubleplus (i' \xmapsto{i} a.m(v)) \doubleplus F', E)}
  {\prd{preemptable}(i', F \doubleplus (i' \xmapsto{i} a.m(v)) \doubleplus F', E)}
\end{equation*}
An invocation is preemtable if its caller has failed (as with cancellation) or
if it is nested in an invocation whose caller has failed (unlike cancellation).
If \prg{a} calls \prg{b} calls \prg{c} and \prg{a} fails, we want to preempt
\prg{b} and \prg{c}. Cancellation on the other hand is designed not to interfere
with running invocations. In this scenario, we cannot cancel \prg{c} even it is
has not started yet because its result is needed to run \prg{b} to completion.

Rule (preempt) removes a runnable, preemptable invocation from the flow even if
this invocation is already running, if so removing the matching process from the
ensemble.

Both, cancellation and preemption proceed from the top of the call stack down
thanks to the runnable precondition. If \prg{a} calls \prg{b} calls \prg{c} and
\prg{a} fails, then \prg{b} cannot be preempted before \prg{c} is. This order
ensures that the chain of happen-before relationships in a call stack is never
broken: \prg{a} waits for \prg{b}, \prg{b} waits for \prg{c}, hence \prg{a}
waits for \prg{c}.

\section{Implementation}
\label{sec:impl}
The message exchanges in the formal semantics closely match the actual
implementation. However the semantics ignores important practical issues such as
how to group actors into operating system processes, how to divide the flow of
messages into separate queues, how to produce and consume messages out of order,
and how to detect failures. We first consider these questions then explain the
reconciliation algorithm that implements fault recovery.

\subsection{Processes and Queues}

Because KAR applications can combine heterogeneous components
implemented using diverse programming languages and middleware
frameworks, we built KAR as an out-of-process runtime. Application
processes can either access the runtime's functionality over HTTP/2 via its
language-neutral REST API
or via SDKs provided for Java, JavaScript, and Python. The entire KAR
implementation is open source~\cite{kar-github-anon}.

A running KAR application consists of a dynamically varying number of
components. Each component consists of a paired application and runtime process.
Each process pair runs on the same logical node (same physical host, virtual
machine, or Kubernetes pod). The (ab)normal termination of one process
triggers the termination of the paired process.

Application processes may be dedicated to specific tasks and replicated for
horizontal scaling. Application processes announce which actor type(s) they
support and KAR automatically places each instantiated actor in a compatible process. The
runtime processes coordinate actor placement using a compare-and-swap operation
on a persistent distributed data store. To reduce the frequency of store
accesses, each runtime process maintains a placement cache that is invalidated
on component failures.

KAR's implementation allocates a dedicated message queue for each application
component. Request messages are added to the callee's queue. Responses are added
to the caller's queue for \prg{actor.call} or to the callee's queue for
\prg{actor.tell}.
A \prg{tell} waits for the queue to acknowledge the message to guarantee
durability. A \prg{call} blocks until a matching response message is received.
In each component, a consumer thread listens for
incoming messages. It delivers responses to suspended callers and
dispatches requests to in-memory per-actor queues, except
for reentrant invocations that bypass the in-memory queues.

KAR's formal semantics makes a convenient but unrealistic assumption
that it is possible to discard or alter messages in the middle of a
queue. Typical production systems do not support this. Therefore, the
implementation (1) only appends a message at the end of a queue and (2)
expires messages in bulk.

Thus, KAR (a) does not dequeue a request message immediately upon completion
and (b) enqueues all tail calls at the end of the callee's
queue. To ensure a tail call to self runs next, KAR does not release the actor
lock and recognizes the tail call as owning the lock (bypassing the
in-memory queue just like a reentrant call).

KAR expires the oldest messages after a
configurable delay or above a configurable queue size. These parameters can
be adjusted on a per-application basis to ensure messages are not expired before
use. By default, messages expire after ten minutes.

\subsection{Kafka and Redis}

We choose Redis~\cite{redis} to coordinate actor placement and implement KAR's
persistence API (\prg{actor.state}).

The runtime processes communicate with one another using Apache
Kafka~\cite{kafka}. Kafka provides not only the reliable message queues, but
also the authentication and discovery mechanisms, the consensus protocol, as
well as the health monitoring and failure detection. Runtime processes
authenticate with Kafka and form a Kafka consumer group for a Kafka topic that
is unique to the application. This topic is dynamically partitioned into
independent queues attached to each runtime process. Kafka detects the removal
of a runtime process using heartbeats. When membership in the consumer
group changes, Kafka allows the list of members to stabilize and
then triggers a recovery, i.e., reconciliation.

Once Kafka removes a runtime process for the consumer group, this process no
longer receives messages from Kafka but it is also prevented from sending more
messages to Kafka, even if the process is not completely dead. Once the consumer
group stabilizes, remaining healthy runtime processes also forcefully disconnect
non-healthy processes from Redis before resuming the execution. In essence, this
algorithm implements the fail-stop failures semantics of~\ref{sec:sem-failures}.
Even if the Kafka failure determination is an approximation, we can guarantee
that an application process deemed to have failed cannot further access or
mutate actor state, hence there can be no overlap between state updates from an
actor deemed failed and a replacement actor. As discussed in
Sections~\ref{sec:intro} and~\ref{sec:model}, a form of forceful disconnection
should be implemented for every stateful service in use by KAR application
components.

\subsection{Reconciliation}

When an application component fails, KAR must decide what to do with
the messages in its queue. All components temporarily stop sending and
receiving messages. They reach a consensus on the list of live
components and elect a reconciliation leader. This leader catalogs
unexpired messages. Requests with a matching response or tail call are
discarded. The remaining requests to failed components have to be
preserved since they have not yet run to completion. KAR invalidates
the placement decisions for all actors that were placed in a failed
component and eagerly chooses a replacement component for those
with pending requests. These pending requests are then appended
to the queues of the selected components. KAR takes this opportunity
to reorder requests messages according to the formal semantics
(by moving tail calls to self earlier). It may be impossible to
replace an actor if no component supports the actor type.
KAR queues requests to unavailable types separately, revisiting this
queue when new components are added to the application.

Request messages for nested calls (\prg{actor.call}) include the request id for the
caller. Reconciliation identifies every request with a pending nested
call by transposing the callee-caller map. It alters the copy of the request
message to include the id of the pending callee. When the request is received,
the presence of this optional callee id instructs the runtime to postpone the
retry of the request until a response from the callee is received, hence
fulfilling happen before.

In the end, queues attached to failed components can be
discarded or flushed for later reuse. A failure during reconciliation simply
restarts reconciliation. Request messages already
copied into the queues of live components are skipped.

To avoid magnifying transient failures, Kafka recommends and
defaults to a 10 second grace period before
deciding a process has failed. Given this time scale,
there is room for a relatively long reconciliation phase.
Section~\ref{sec:eval} reports that in our experiments,
slightly less than half of the total
detection and recovery time was spent on reconciliation.

Reconciliation time increases with the number of recent messages hence
application components. So for larger scale systems, a different implementation
may be necessary. An alternative to reconciliation could use Kafka
transactions~\cite{kafka-transactions} to atomically (1)~send the caller the
call result via the caller's queue and (2) log its completion in the callee's
queue, making it possible to match requests and completions within each failed
component queue without global coordination. We leave this for future work.

\subsection{Cancellation}

The formal semantics permit callees to be optionally preempted or
canceled when the callers fail. Pragmatically, we chose to
not implement preemption, as terminating running tasks can easily lead
to an actor's state becoming inconsistent. To efficiently implement
cancellation, just before executing the callee its runtime process
checks the list of live application components established during
the most recent reconciliation. If the caller's component is not listed,
execution is elided and a synthetic response message sent.

\section{Container Shipping Application}
\label{sec:app}
To demonstrate how KAR could be applied to a typical enterprise
scenario, we teamed with a team of client-facing cloud developers to
explore the suitability
of KAR for their pre-existing Container Shipping application.
The application models a subset of the business processes of a maritime
shipping company. Clients can place orders to arrange shipping on a
specific ship voyage of temperature sensitive products which require
one or more refrigerated containers. Voyages are assigned to a shipping
schedule between ports. Ships depart and arrive as scheduled
and periodically broadcast their positions while in transit.  At any
time a container can suffer an anomaly indicating a failure of
refrigeration. The detection of non-functional containers triggers
different business logic depending on their state: in-transit, assigned to an order
before departure, or empty.

This application was originally built without KAR. The development team designed it both to capture a specific
customer scenario and to serve as a reference application for
event-driven architectures. Interestingly, the original
design~\cite{icg-shipping-arch} was described in terms of actors and
their interactions, but the actual implementation was a collection of
classic microservices connected via a Kafka-based event
bus~\cite{icg-shipping-impl}. In the process of deriving an efficient
implementation, the natural actor-based domain model
was abandoned. As a demonstration of KAR's capabilities, we
re-implemented the core business logic following the original
actor-based design in approximately 5,000 lines of Java code using the
OpenLiberty-based KAR Java SDK.

\begin{figure*}[t]
  \centering
  \vspace{-0.1in}
  \begin{subfigure}{0.48 \textwidth}
    \includegraphics[trim=40 40 120 0,clip,width=\textwidth]{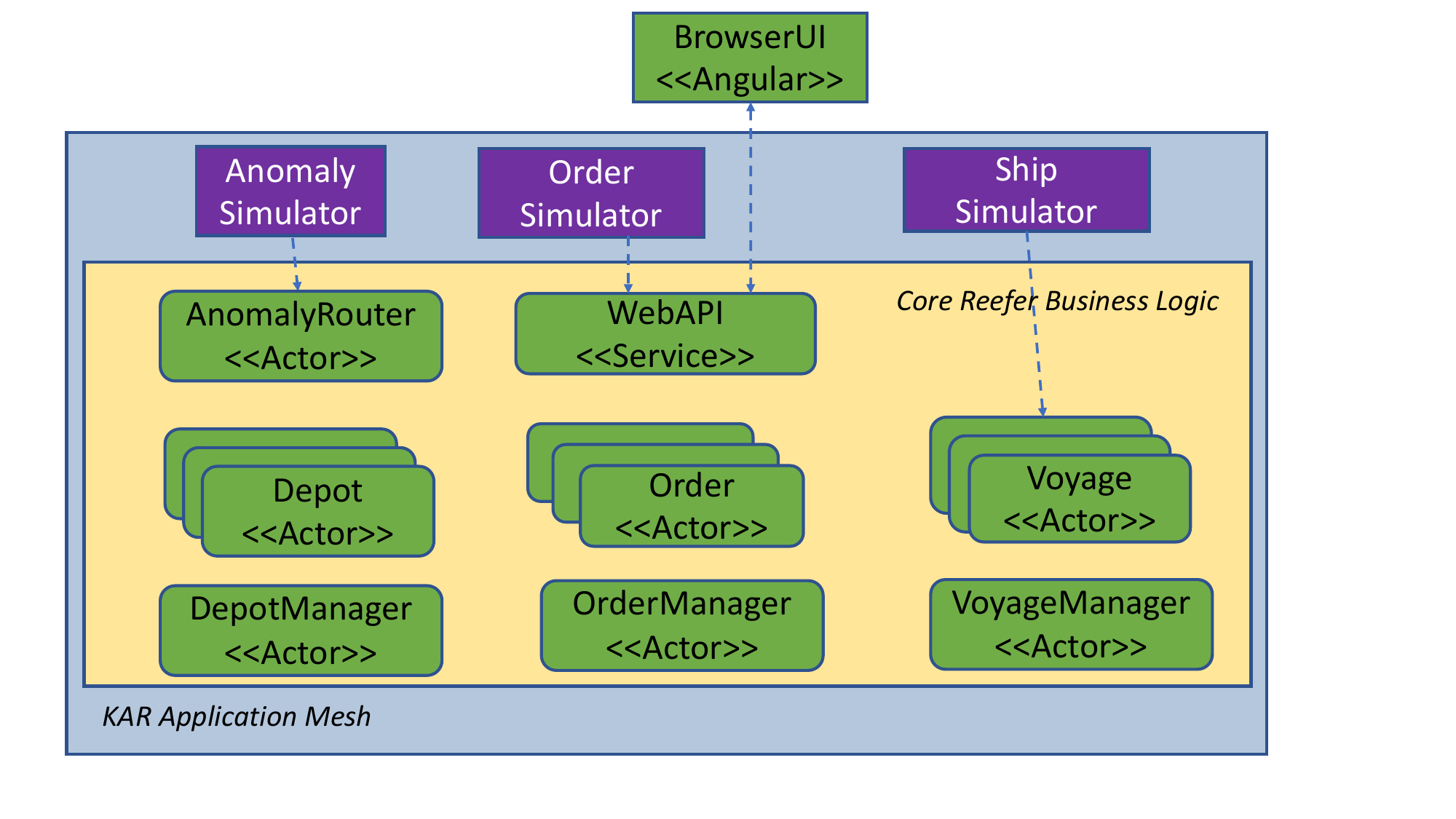}
      \caption{Logical application architecture}
      \Description{Container Shipping application software architecture}
      \label{fig:shipping-architecture}
  \end{subfigure}
  \begin{subfigure}{0.48 \textwidth}
    \includegraphics[trim=85 10 70 0,clip,clip,width=\textwidth]{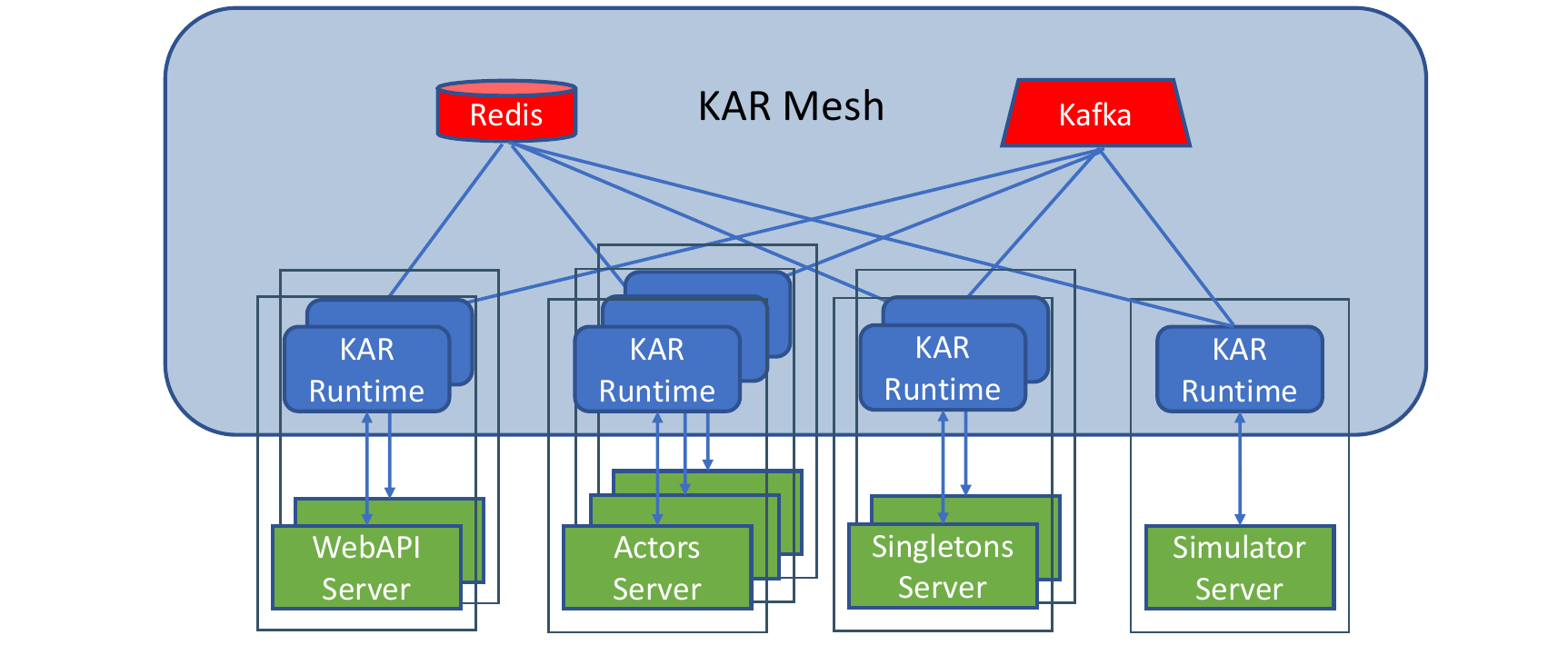}
    \caption{Replicated application deployment.}
    \label{fig:shipping-deployment}
  \Description{Replicated application deployment}
  \end{subfigure}
  \vspace{-0.1in}
  \caption{Software architecture and runtime deployment of Container Shipping Application}
\end{figure*}

The high level software components of the application are shown in
\autoref{fig:shipping-architecture}.
The \prg{BrowserUI} is implemented with Angular and is used to
visualize application behavior and provide an easy interface for
controlling the simulators that drive it.
The \prg{WebAPI} is a stateless service that provides the application interface.
It receives
updates from stateful actor components and pushes that information in real
time to \prg{BrowserUI}.
The \prg{Order} actor implements order logic and maintains persistent state for
a single order. An order actor instance is instantiated on an order creation
request and its state is removed upon arrival at its destination port.
The \prg{Voyage} actor implements voyage logic and maintains the persistent state
for a single voyage. A voyage actor instance is instantiated on its first order
or on departure if empty. Its state is removed upon arrival.
The \prg{Depot} actor implements container management logic and manages the
container inventory for a port. Container data includes
references to its owning order and voyage.
The \prg{AnomalyRouter} is a singleton actor that maintains a mapping
of container locations that enables it to route container anomaly events to
the appropriate depot or voyage actor. There are also
singleton \prg{OrderManager}, \prg{VoyageManager},
and \prg{DepotManager} actors that manage global state and track
statistics. 
\autoref{fig:shipping-deployment} illustrates the intended production deployment
of the application with replicated instances of each component to support
failover and scale out. The \prg{Order}, \prg{Voyage}, and \prg{Depot}
actor types are configured to be hosted on the ``Actors Server'' while the
other actor types are hosted on the ``Singletons Server''; this enables
these two actor-hosting components to be scaled independently. 

Container application components are driven by events including: order creation
requests, voyage position changes, and container anomalies. Custom event
simulators have been developed to automatically generate load to stress both the
application and KAR. The simulators maintain no application state. They
interface with the WebAPI and relevant actors.
Simulated voyage fill target, order generation rate, and anomaly rate
are controlled through \prg{BrowserUI}.

\begin{figure}[t]
  \vspace{-0.1in}
  \centering
  \includegraphics[width=0.9 \textwidth]{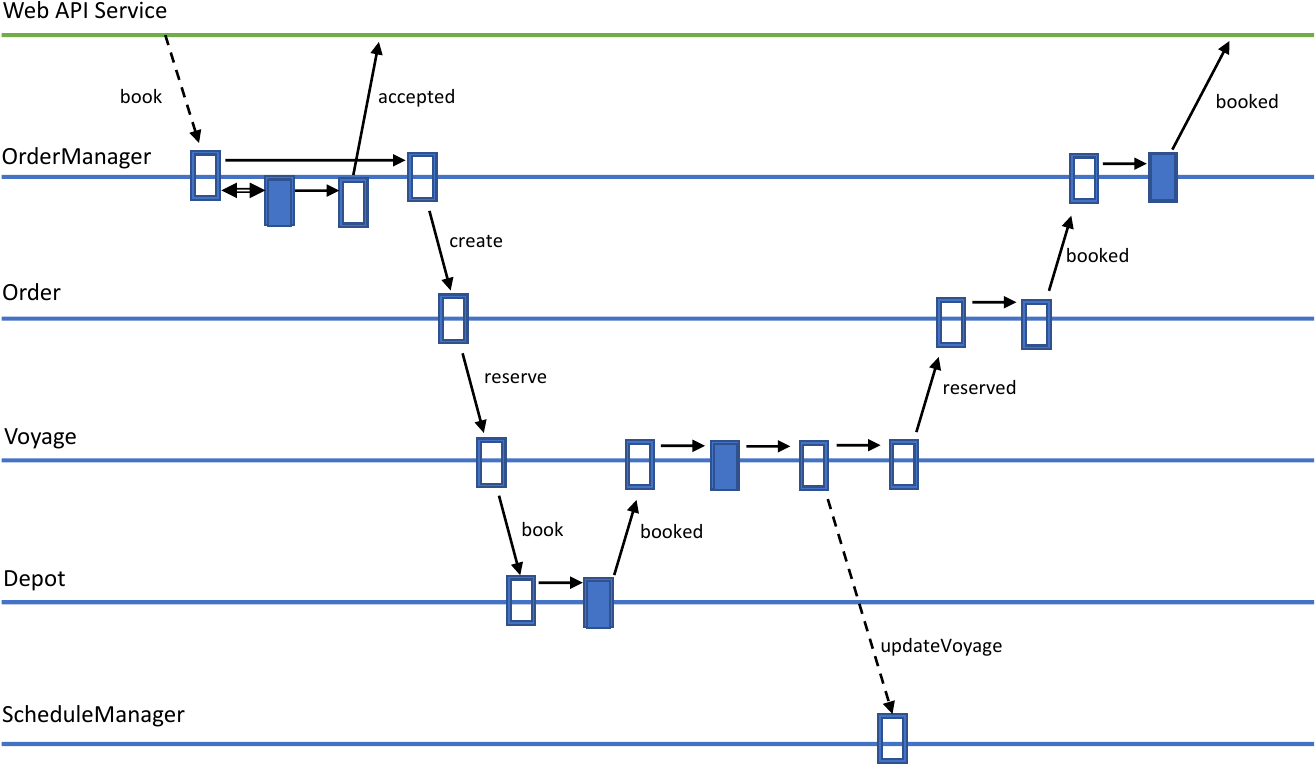}
  \vspace{-0.1in}
  \caption{Depiction of the workflow for accepting and processing a client order in the Container Shipping
  application. Each box represents an actor method invocation. Shaded
  boxes indicate invocations that update external state. Single-headed
  arrows with solid lines indicate tail calls. Double-headed arrows
  indicate synchronous calls. Single-headed arrows with dotted lines
  indicate asynchronous tells.}
  \Description{Process interaction diagram depicting the interactions between the
  five actor instances involved in the business process of accepting an order
  from a client.}
  \label{fig:orderflow}
\end{figure}

To illustrate how the implementation of the application's business
logic benefits from KAR's capabilities as summarized in \autoref{sec:benefits},
\autoref{fig:orderflow} depicts the steps taken during a successful
execution of the workflow for accepting and processing a client order.
The execution spans 17 methods of 5 actor types, and is primarily
orchestrated by tail calls. However it also includes the use of a
synchronous reentrant call by the \prg{Order} actor to create a
sub-orchestration that ends by notifying the WebAPI that the order has
been accepted and an asynchronous tell that spawns a background update
of the \prg{ScheduleManager}. Four of the methods, indicated by shaded
boxes, update externally
managed state; tail calls isolate these updates to simplify the
programmer's reasoning about retrying them on failures. Throughout the
workflow, failure recovery is simplified by the purely local reasoning
enabled by KAR's retry orchestration.

\section{Evaluation}
\label{sec:eval}
Our empirical evaluation is designed to provide evidence that supports
two key hypotheses.
\begin{enumerate}
\item KAR-based applications can automatically and reliably detect and
recovery from failures.
\item KAR's actor-based programming model and external runtime system
are not significantly less performant than the reliable messaging
system on which they are built.
\end{enumerate}

\subsection{Failure Detection and Recovery}
\label{sec:eval:faults}

In KAR's target environment of the cloud, failures are relatively
infrequent, but may be clustered. As a result of a failure, one or
more application components may exit abruptly and without
notification. We designed our fault-testing scenarios accordingly,
with the time-saving modification of ``fast forwarding'' through the
failure-free intervals by randomly injecting the next fault less than
two minutes after the application fully recovers from the previous one.

We use the Container Shipping application described in the previous
section as the user workload for our fault-injection experiments.
To perform the experiments in a completely controlled fashion, we use
\prg{k3d}~\cite{k3d-website} to create a virtual five node Kubernetes
cluster. Using node affinities, we deploy all Kubernetes system pods,
Kafka, Redis, and the simulators on three nodes of this cluster. We
deploy two replicas of each remaining application component on the
other two ``victim'' nodes.
Using an automated test harness, we cause a series of abrupt
failures by instructing \prg{k3d} to do a hard stop on a randomly
selected victim node. Each failure results in the abrupt
termination of multiple application and runtime processes.
After normal application operation has automatically recovered on replica
processes, the node is restarted which spawns new replicas.

\begin{figure*}[t]
  \centering
  \vspace{-0.1in}
  \begin{subfigure}{0.48 \textwidth}
    \includegraphics[width=\textwidth]{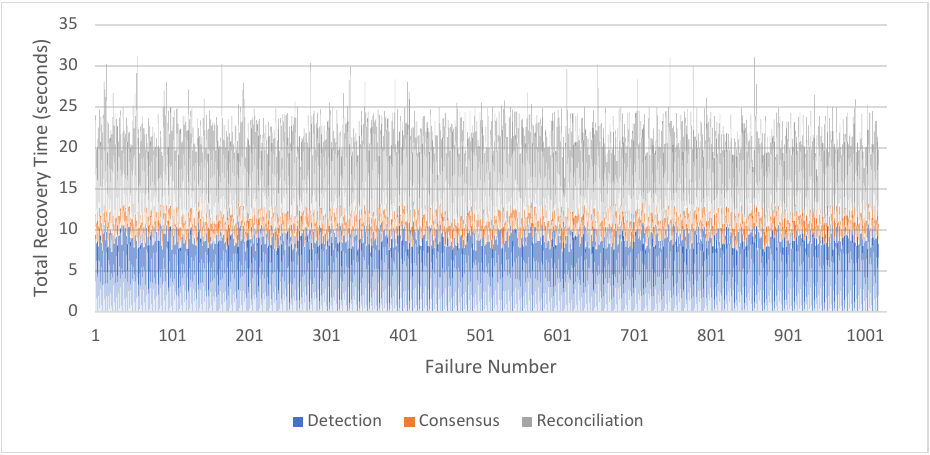}
    \caption{Phases of failure detection and recovery.}
    \Description{A graph showing the time in seconds taken by each major phase of fault recovery in KAR}
    \label{fig:shipping-recovery}
  \end{subfigure}
  \begin{subfigure}{0.48 \textwidth}
    \includegraphics[width=\textwidth]{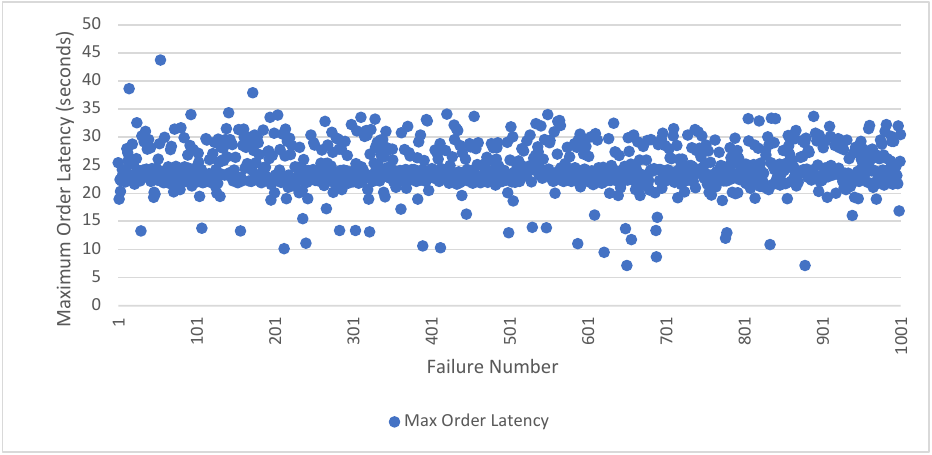}
    \caption{Maximum order latency around failure time.}
    \label{fig:shipping-order-latency}
  \Description{A graph showing the maximum order latency in seconds 
  during a time period in which a failure and recovery is occurring}
  \end{subfigure}
  \vspace{-0.1in}
  \caption{Analysis of 1,000 single node failures during a 48
  hour run of the Container Shipping application.}
\end{figure*}

Since the node running the simulators is never killed, we can easily
verify that failures never cause submitted orders to be lost. The
application also dynamically checks additional application-level invariants,
such as that ships arrive and depart as scheduled carrying
their expected cargoes, that ships and containers neither disappear or
appear out of thin air, and that simulation time continuously
advances. None of these invariants were ever violated during the
experiments.

During a 48 hour run of the application, we injected 1,000 single node
failures. The typical outage caused by a failure lasted for 22
seconds, with a minimum elapsed time of 16.1 seconds and a maximum of
31.2 seconds. The typical failure took 9 seconds to detect followed by
an additional 13 seconds to recover before normal execution resumed.
\autoref{fig:shipping-recovery} breaks each such outage into
its three main components: the time for Kafka to detect a
failure, the time to reach a distributed consensus on the new
application topology, and the time spent by KAR in
reconciliation. \autoref{tab:shippingstats} reports the detailed
statistics. On average, reconciliation time is just under half of the
total outage time.

\begin{table}[t]
\centering
\caption{Summary statistics for 1,000 failures (seconds)}
\label{tab:shippingstats}
\small
\begin{tabular}{l|r|r|r|r|r}
  \toprule
               & Average & StdDev & Median & Min & Max  \\
  \midrule
  Total Outage & 22.139	& 2.114 & 22.015 & 16.117 & 31.207 \\
  \midrule
  \midrule
  Detection    & 9.053	& 0.907 & 9.084 & 7.217	& 11.022 \\
  \midrule
  Consensus  &  2.437 & 0.086 & 2.443 & 2.232	& 3.197 \\
  \midrule
  Reconciliation & 10.649 & 1.967 & 9.098 & 6.019 & 21.035 \\
  \bottomrule
\end{tabular}
\vspace{-0.2in}
\end{table}

\autoref{fig:shipping-order-latency} presents an application-level
view of the outages by showing a scatter plot of the maximum order
latency observed in each time window surrounding a failure. In
failure-free operation, the average order latency is around 100
milliseconds. Around failures, this spikes to an average (median) of
24.5 (24.0) seconds, with a maximum of 43.8 seconds and a minimum of
7.2 seconds. The maximum order latency during a failure is
occasionally less than the outage time because the application is
replicated. During the handful of failures where all in-flight orders
were being handled by unimpacted replicas, processing could continue
normally until interrupted by the consensus and reconciliation phases.

We also tested two more challenging failure scenarios.  First we
verified that KAR can robustly handle failures during recovery by
injecting 1,000 paired node failures where the second failure was
timed to occur during the consensus or reconciliation phases of
recovery.  Second, we performed 500 iterations of a complete
application failure scenario where all application and runtime
processes except the simulator were killed abruptly and then restarted
after waiting for 30 seconds. KAR and the application were able to
handle all of these failures successfully.

\subsection{Reliable Messaging}
\label{sec:eval:messaging}

Our experimental testbed is a five node Kubernetes v1.22 cluster
provisioned via the managed Kubernetes service of the IBM public
cloud. Each (virtual) worker node is a \prg{b3c.4x16} configuration,
which has 4 CPUs, 16 GB of memory, and runs Ubuntu 18.04. The nodes are connected via a 1Gbps
virtual private network. We deploy KAR v1.3.1 on this cluster in conjunction
with three different Kafka and Redis configurations:
{\em ClusterDev}, {\em ClusterProd}, and {\em Managed}.
The {\em Cluster} configurations run deployments of Kafka and Redis
within the Kubernetes cluster.
With {\em ClusterDev}, Kafka and Redis data is not
backed by persistent storage and there is only one replica of each.
With {\em ClusterProd}, Kafka and Redis data is backed by
attached Persistent Volumes that support 1000 IOPS;
Kafka is configured with 3-way replication.
With {\em Managed}, Kafka and Redis
are instantiated using IBM's fully managed production services:
Event Streams and Databases for Redis. These managed instances are
provisioned in the same cloud region as the Kubernetes cluster.

Table~\ref{tab:latency} reports the end-to-end latency in milliseconds
of a minimal request-response communication pattern with a small
payload (20 bytes of user data) on these three different KAR system
configurations. We report the median latency of 10,000 iterations. The
two communicating processes are placed on different worker
nodes. The first two columns are baseline measurements that do not
involve the KAR runtime. The {\em Direct HTTP}
column reports the time required for a non-reliable request/response
communication (a {\tt POST} request over an HTTP connection)
between two Node.js processes. The second column, {\em Kafka Only}
isolates the end-to-end latency for two Go processes that send
messages by connecting directly to Kafka using the same Go-based
client as the KAR runtime. The final two columns report the latency
for KAR actor method invocations using the Node.js KAR SDK. {\em Kar
Actor} is the default configuration; {\em KAR Actor (no cache)}
disables the actor placement cache that short circuits
Redis access on most actor method invocations.

\begin{table}[ht]
\centering
\caption{Median round trip message latency in milliseconds}
\label{tab:latency}
  \begin{tabular}{l|r|r|r|r}
  \toprule
               & Direct & Kafka & KAR Actor & KAR Actor  \\
               & HTTP   & Only  &           & (no cache) \\
  \midrule
  ClusterDev   &  2.60  &  4.35 &    6.62   &    7.12    \\
  \midrule
  ClusterProd  &  2.60  & 10.62 &   13.41   &   14.31    \\
  \midrule
  Managed      &  2.60  & 14.56 &   15.80   &   18.06    \\
  \bottomrule
  \end{tabular}
\end{table}

The first conclusion from these numbers is that, unsurprisingly, there
is a measurable cost to communicating through reliable message queues
{\em vs.}  non-resilient direct HTTP connections. When Kafka is
replicated for fault tolerance, the {\em Kafka Only} request-response
latency is $4X$ to $5.6X$ higher than in {\em Direct HTTP}. These
Kafka latencies are comparable with others reported in the literature,
for example \citet{RSM:ecoop2019} reports an end-to-end message
latency of 9 milliseconds using a replicated cloud-deployed Kafka
configuration.  Second, the extra inter-process communication
introduced by the KAR external runtime design and the bookkeeping
associated with actor method invocation only adds modest overhead to
the base cost of using reliable messaging. In the configurations that
we believe are representative of KAR's intended production use cases,
{\em Managed} and {\em ClusterProd}, {\em KAR Actor} incurs less than
$20\%$ additional latency when compared to {\em Kafka Only}. Finally,
we can see that although caching actor placement only provides
a marginal benefit in {\em ClusterDev} and {\em ClusterProd}, it has
more impact in {\em Managed} where the Redis instance is not
co-located in the same Kubernetes cluster as the application
processes.

\section{Related Work}
\label{sec:related}
Our programming model and some aspects of its implementation were
inspired by the {\em Distributed Application Runtime} (DAPR)
project~\cite{dapr-website,dapr-github}. Both systems delegate dynamic
service discovery and cross-component communication to a mesh of
external runtime processes. The runtime exposes a language neutral
REST API, but is augmented with SDKs for selected
languages and middleware frameworks to simplify programming.

The virtual actor model that is used by both DAPR and KAR was
one of the main innovations of the Orleans
system~\cite{Bykov:2011:OCC:2038916.2038932,bernstein2014orleans}.
Virtual actors improve on the usability of previous actor systems such
as Akka~\cite{akka} and Erlang~\cite{erlang:cacm} by making actor
placement and life-cycle management the responsibility of the runtime
system instead of the application programmer.

The application-level fault tolerance capabilities enabled by KAR go
beyond those provided by DAPR or Orleans.  Both previous systems view
individual actors as the unit of fault tolerance and recovery, and
make weaker guarantees about the failure semantics of
multi-actor interactions. DAPR and Orleans offer reliable,
at-least-once message delivery, but do not guarantee that a
successfully completed message will never be re-delivered as part of
failure recovery. KAR is the only one of the three systems that fully
explores the combination of call chain actor reentrancy and fault
tolerance. Orleans 2.x did not correctly implement full call chain
reentrancy~\cite{orleans-call-chain-bug}, and call chain reentrancy
has been removed in later versions~\cite{orleans-call-chain-removed}.
DAPR v1.6 provides call chain reentrancy as a preview feature that is
not supported by any of its SDKs and does not describe how enabling it
will interact with failure recovery~\cite{dapr-call-chain}.

Reliable State Machines (RSM) is a framework for
developing cloud-native applications with a strong emphasis on
fault-tolerance. \citet{RSM:ecoop2019} defines the formal semantics of
RSM and describes an implementation built on
reliable messaging services provided by Microsoft Azure.
Like RSM, KAR relies on reliable message queues to connect application
components and build a replay-able log, but unlike RSM, KAR permits
applications to persist state outside of this log.  KAR
extends on the RSM programming model by making state machines and
actor instances orthogonal concepts. KAR state transitions use
the same tail call mechanism within and across actors.
Furthermore, KAR
supports composing state machines with other programming
patterns for achieving fault tolerance.

The Resilient X10 system~\cite{Res-X10:TOPLAS,Crafa2014:ECOOP} emphasizes the
importance of preserving the {\em happens-before invariant} for enabling
fault-tolerant distributed programming. X10's finish-async programming
model supports a more general task graph structure than KAR's simple
caller-callee relationship. At the time of a failure, a KAR task can
have at most one live child; X10 tasks may have many.
As a result, X10 requires substantially more complicated
fine-grained distributed bookkeeping to ensure that a task cannot
finish, hence be retried, before any of its subtasks.
Unlike KAR, Resilient X10 does not automatically
retry failed tasks. KAR automates and orchestrates retries, thus
eliminating much of the hand-written application-specific recovery
code required by X10.

Many prior systems simply retry apparently failed tasks to achieve a
measure of fault
tolerance~\cite{Dean:2004,hadoop,White:2009:HDG:1717298,Zaharia2012:USENIX}.
This works well with side-effect free tasks, but is
challenging to apply effectively to complex workflows that can contain
non-idempotent operations or interactions with external systems.

HPC applications have long relied on coordinated checkpoint/restart
both as a mechanism for resiliency and to decompose long-running
applications into more schedulable units of
work~\cite{Elnozahy02acm,Sato12sc}. Checkpoint/restart has also been
explored for fault tolerance in distributed actor
systems. \citet{Field:POPL2005} develop a formal semantics of
transactors and show that it is possible to obtain globally consistent
checkpoints in a loosely coupled system by exploiting the structure of
actor interactions. The AEON system~\cite{Sang:OOPSLA2020} explores a
practical implementation of this approach. AEON restricts
actor interactions to occurring in directed acylic graphs (DAG) and
exploits their dominator structure to drive checkpointing
operations. In contrast, KAR's support for fault-tolerant reentrancy
avoids restricting actor interactions to DAGs and enables cyclic
multi-actor interactions.

Transaction protocols like two phase commit provide fault tolerance over grouped
operations for online transactional processing and ideally provide ACID
guarantees~\cite{10.1145/1132863.1132867}.
Later transaction protocols target distributed architectures and minimize
locking to achieve higher scalability and performance but with weaker
guarantees.  For example, SAGA~\cite{10.1145/38713.38742} can abort a
transaction using sequentially executed, compensatory actions to rollback
independently committed operations to a clean state, while Thorp optimizes
2PL/2PC to support actor-based transactions in the Microsoft Orleans
service~\cite{eldeeb2016transactions}. Beldi~\cite{258880} extends
Olive~\cite{199350} and utilizes transactions to provide exactly once
semantics to ``stateful serverless functions'' -- serverless functions
that maintain state in external NoSQL databases. Beldi accomplishes this
by providing a library that wraps select NoSQL stores; all access by
the functions to the NoSQL store must go through Beldi's API to
realize the the promised exactly once execution semantics.

Durable Functions extends Azure Functions~\cite{microsoft-azure-functions}
with entities and orchestrations whose state/progress is automatically persisted
and restored after a failure~\cite{durable-functions}.
\citet{Burckhardt:OOPSLA21} formalizes an idealized failure-free semantics for
Durable Functions, and establishes that even in the presence of failures a {\em
compute-storage} model preserves an {\em observably exactly-once} execution of
the Durable Functions application. Key to this proof is the assumption that all observable
application state (entity state, message queues, and orchestration progress) is
persisted in a single durable store that is managed by the Durable Functions runtime and can be
updated atomically. Enabled by this strong assumption about application state,
Durable Functions offers a weaker retry semantics than KAR. Durable Functions allows multiple
concurrent executions of the same work item (only one of which will be able to
successfully commit its updates to the store on completion)
(\cite{Burckhardt:OOPSLA21} section 5.3).
Unlike Durable Functions, KAR embraces an open world assumption.
KAR applications are not
restricted to using a single system managed store. KAR's more precise retry
orchestration facilitates interactions with external services that may have
irrevocable side effects.
KAR also provides a more flexible
programming model by not imposing a strict stratification between activity
functions, entity functions, and deterministic orchestrator functions.
KAR actors combine mutable state and control-flow state (active tail
call in a chain).
Moreover, KAR actor methods need not be deterministic, making it easy
for instance for a task to do something else after repeated retries.

\section{Conclusions}
\label{sec:conc}
An increasingly rich collection of applications are
being built for and deployed on diverse cloud platforms. The cloud is
no longer just a platform for stateless functions, batch analytics, or
other fault-oblivious side-effect free computations. Enterprises are
migrating mission-critical stateful applications that interact with multiple
external stateful services to the cloud. KAR provides a
programming model and supporting runtime that is designed to simplify
and support the development of such applications, while taking
advantage of existing cloud capabilities.  In the
presence of failures, KAR not only orchestrates retries of individual
tasks but also worries about verticals and horizontals, i.e, call
stacks and call chains.
KAR demonstrates it is possible to productively develop fault-tolerant
applications even if the application state is split
among multiple, independent persistent services.
In this paper, we formalized the core semantics of KAR to precisely
formulate and establish its fault tolerance guarantees and outlined
an implementation built on Kafka.


\bibliography{main}


\begin{thebibliography}{38}


\ifx \showCODEN    \undefined \def \showCODEN     #1{\unskip}     \fi
\ifx \showDOI      \undefined \def \showDOI       #1{#1}\fi
\ifx \showISBNx    \undefined \def \showISBNx     #1{\unskip}     \fi
\ifx \showISBNxiii \undefined \def \showISBNxiii  #1{\unskip}     \fi
\ifx \showISSN     \undefined \def \showISSN      #1{\unskip}     \fi
\ifx \showLCCN     \undefined \def \showLCCN      #1{\unskip}     \fi
\ifx \shownote     \undefined \def \shownote      #1{#1}          \fi
\ifx \showarticletitle \undefined \def \showarticletitle #1{#1}   \fi
\ifx \showURL      \undefined \def \showURL       {\relax}        \fi
\providecommand\bibfield[2]{#2}
\providecommand\bibinfo[2]{#2}
\providecommand\natexlab[1]{#1}
\providecommand\showeprint[2][]{arXiv:#2}

\bibitem[Akka(2011)]%
        {akka}
Akka \bibinfo{year}{2011}\natexlab{}.
\newblock \bibinfo{title}{Akka Actor Model}.
\newblock \bibinfo{howpublished}{\url{https://akka.io/docs/}}.
\newblock


\bibitem[Apache(2016)]%
        {kafka-transactions}
\bibfield{author}{\bibinfo{person}{Apache}.} \bibinfo{year}{2016}\natexlab{}.
\newblock \bibinfo{title}{KIP-98 - Exactly Once Delivery and Transactional
  Messaging}.
\newblock
  \bibinfo{howpublished}{\url{https://cwiki.apache.org/confluence/display/KAFKA/KIP-98+-+Exactly+Once+Delivery+and+Transactional+Messaging}}.
\newblock


\bibitem[Armstrong(2010)]%
        {erlang:cacm}
\bibfield{author}{\bibinfo{person}{Joe Armstrong}.}
  \bibinfo{year}{2010}\natexlab{}.
\newblock \showarticletitle{Erlang}.
\newblock \bibinfo{journal}{\emph{Commun. ACM}} \bibinfo{volume}{53},
  \bibinfo{number}{9} (\bibinfo{date}{Sept.} \bibinfo{year}{2010}),
  \bibinfo{pages}{68–75}.
\newblock
\showISSN{0001-0782}
\urldef\tempurl%
\url{https://doi.org/10.1145/1810891.1810910}
\showDOI{\tempurl}


\bibitem[Bernstein et~al\mbox{.}(2014)]%
        {bernstein2014orleans}
\bibfield{author}{\bibinfo{person}{Phil Bernstein}, \bibinfo{person}{Sergey
  Bykov}, \bibinfo{person}{Alan Geller}, \bibinfo{person}{Gabriel Kliot}, {and}
  \bibinfo{person}{Jorgen Thelin}.} \bibinfo{year}{2014}\natexlab{}.
\newblock \bibinfo{booktitle}{\emph{Orleans: Distributed Virtual Actors for
  Programmability and Scalability}}.
\newblock \bibinfo{type}{{T}echnical {R}eport} MSR-TR-2014-41.
\newblock
\urldef\tempurl%
\url{https://www.microsoft.com/en-us/research/publication/orleans-distributed-virtual-actors-for-programmability-and-scalability/}
\showURL{%
\tempurl}


\bibitem[Burckhardt et~al\mbox{.}(2021)]%
        {Burckhardt:OOPSLA21}
\bibfield{author}{\bibinfo{person}{Sebastian Burckhardt},
  \bibinfo{person}{Chris Gillum}, \bibinfo{person}{David Justo},
  \bibinfo{person}{Konstantinos Kallas}, \bibinfo{person}{Connor McMahon},
  {and} \bibinfo{person}{Christopher~S. Meiklejohn}.}
  \bibinfo{year}{2021}\natexlab{}.
\newblock \showarticletitle{Durable Functions: Semantics for Stateful
  Serverless}.
\newblock \bibinfo{journal}{\emph{Proc. ACM Program. Lang.}}
  \bibinfo{volume}{5}, \bibinfo{number}{OOPSLA}, Article
  \bibinfo{articleno}{133} (\bibinfo{date}{Oct.} \bibinfo{year}{2021}),
  \bibinfo{numpages}{27}~pages.
\newblock
\urldef\tempurl%
\url{https://doi.org/10.1145/3485510}
\showDOI{\tempurl}


\bibitem[Bykov et~al\mbox{.}(2011)]%
        {Bykov:2011:OCC:2038916.2038932}
\bibfield{author}{\bibinfo{person}{Sergey Bykov}, \bibinfo{person}{Alan
  Geller}, \bibinfo{person}{Gabriel Kliot}, \bibinfo{person}{James~R. Larus},
  \bibinfo{person}{Ravi Pandya}, {and} \bibinfo{person}{Jorgen Thelin}.}
  \bibinfo{year}{2011}\natexlab{}.
\newblock \showarticletitle{Orleans: cloud computing for everyone}. In
  \bibinfo{booktitle}{\emph{Proc. 2nd ACM Symposium on Cloud Computing}}
  (Cascais, Portugal) \emph{(\bibinfo{series}{SOCC '11})}.
  \bibinfo{publisher}{ACM}, \bibinfo{address}{New York, NY, USA}, Article
  \bibinfo{articleno}{16}, \bibinfo{numpages}{14}~pages.
\newblock
\showISBNx{978-1-4503-0976-9}
\urldef\tempurl%
\url{https://doi.org/10.1145/2038916.2038932}
\showDOI{\tempurl}


\bibitem[Crafa et~al\mbox{.}(2014)]%
        {Crafa2014:ECOOP}
\bibfield{author}{\bibinfo{person}{Silvia Crafa}, \bibinfo{person}{David
  Cunningham}, \bibinfo{person}{Vijay Saraswat}, \bibinfo{person}{Avraham
  Shinnar}, {and} \bibinfo{person}{Olivier Tardieu}.}
  \bibinfo{year}{2014}\natexlab{}.
\newblock \showarticletitle{Semantics of ({Resilient}) {X10}}.
\newblock In \bibinfo{booktitle}{\emph{Proc. 28th European Conference on
  Object-Oriented Programming}}. \bibinfo{pages}{670--696}.
\newblock
\urldef\tempurl%
\url{https://doi.org/10.1007/978-3-662-44202-9_27}
\showDOI{\tempurl}


\bibitem[Cutting and Baldeschwieler(2007)]%
        {hadoop}
\bibfield{author}{\bibinfo{person}{Doug Cutting} {and} \bibinfo{person}{Eric
  Baldeschwieler}.} \bibinfo{year}{2007}\natexlab{}.
\newblock \showarticletitle{Meet {H}adoop}. In
  \bibinfo{booktitle}{\emph{{O'Reilly Open Software Convention}}}.
  \bibinfo{address}{Portland, OR}.
\newblock


\bibitem[DAPR(2020a)]%
        {dapr-github}
DAPR \bibinfo{year}{2020}\natexlab{a}.
\newblock \bibinfo{title}{DAPR GitHub Organization}.
\newblock \bibinfo{howpublished}{\url{https://github.com/dapr}}.
\newblock


\bibitem[DAPR(2020b)]%
        {dapr-website}
DAPR \bibinfo{year}{2020}\natexlab{b}.
\newblock \bibinfo{title}{DAPR Project Website}.
\newblock \bibinfo{howpublished}{\url{https://dapr.io}}.
\newblock


\bibitem[DAPR Reentrancy(2022)]%
        {dapr-call-chain}
DAPR Reentrancy \bibinfo{year}{2022}\natexlab{}.
\newblock \bibinfo{title}{How-to: Enable and use actor reentrancy in Dapr}.
\newblock
  \bibinfo{howpublished}{\url{https://docs.dapr.io/developing-applications/building-blocks/actors/actor-reentrancy/}}.
\newblock


\bibitem[Dean and Ghemawat(2004)]%
        {Dean:2004}
\bibfield{author}{\bibinfo{person}{Jeffrey Dean} {and} \bibinfo{person}{Sanjay
  Ghemawat}.} \bibinfo{year}{2004}\natexlab{}.
\newblock \showarticletitle{{MapReduce}: Simplified data processing on large
  clusters}. In \bibinfo{booktitle}{\emph{Proc. 6th Conference on Symposium on
  Operating Systems Design \& Implementation}} (San Francisco, CA)
  \emph{(\bibinfo{series}{OSDI'04})}. \bibinfo{pages}{10--10}.
\newblock


\bibitem[Eldeeb and Bernstein(2016)]%
        {eldeeb2016transactions}
\bibfield{author}{\bibinfo{person}{Tamer Eldeeb} {and} \bibinfo{person}{Phil
  Bernstein}.} \bibinfo{year}{2016}\natexlab{}.
\newblock \bibinfo{booktitle}{\emph{Transactions for Distributed Actors in the
  Cloud}}.
\newblock \bibinfo{type}{{T}echnical {R}eport} MSR-TR-2016-1001.
\newblock
\urldef\tempurl%
\url{https://www.microsoft.com/en-us/research/publication/transactions-distributed-actors-cloud-2/}
\showURL{%
\tempurl}


\bibitem[Elnozahy et~al\mbox{.}(2002)]%
        {Elnozahy02acm}
\bibfield{author}{\bibinfo{person}{E.~N. Elnozahy}, \bibinfo{person}{Lorenzo
  Alvisi}, \bibinfo{person}{Yi-Min Wang}, {and} \bibinfo{person}{David~B.
  Johnson}.} \bibinfo{year}{2002}\natexlab{}.
\newblock \showarticletitle{{A survey of rollback-recovery protocols in
  message-passing systems}}.
\newblock \bibinfo{journal}{\emph{ACM Computing Survey}} \bibinfo{volume}{34},
  \bibinfo{number}{3} (\bibinfo{year}{2002}), \bibinfo{pages}{375--408}.
\newblock


\bibitem[Field and Varela(2005)]%
        {Field:POPL2005}
\bibfield{author}{\bibinfo{person}{John Field} {and} \bibinfo{person}{Carlos~A.
  Varela}.} \bibinfo{year}{2005}\natexlab{}.
\newblock \showarticletitle{Transactors: A Programming Model for Maintaining
  Globally Consistent Distributed State in Unreliable Environments}. In
  \bibinfo{booktitle}{\emph{Proceedings of the 32nd ACM SIGPLAN-SIGACT
  Symposium on Principles of Programming Languages}} (Long Beach, California,
  USA) \emph{(\bibinfo{series}{POPL '05})}. \bibinfo{publisher}{Association for
  Computing Machinery}, \bibinfo{address}{New York, NY, USA},
  \bibinfo{pages}{195–208}.
\newblock
\showISBNx{158113830X}
\urldef\tempurl%
\url{https://doi.org/10.1145/1040305.1040322}
\showDOI{\tempurl}


\bibitem[Garcia-Molina and Salem(1987)]%
        {10.1145/38713.38742}
\bibfield{author}{\bibinfo{person}{Hector Garcia-Molina} {and}
  \bibinfo{person}{Kenneth Salem}.} \bibinfo{year}{1987}\natexlab{}.
\newblock \showarticletitle{Sagas}. In \bibinfo{booktitle}{\emph{Proceedings of
  the 1987 ACM SIGMOD International Conference on Management of Data}} (San
  Francisco, California, USA) \emph{(\bibinfo{series}{SIGMOD '87})}.
  \bibinfo{publisher}{Association for Computing Machinery},
  \bibinfo{address}{New York, NY, USA}, \bibinfo{pages}{249–259}.
\newblock
\showISBNx{0897912365}
\urldef\tempurl%
\url{https://doi.org/10.1145/38713.38742}
\showDOI{\tempurl}


\bibitem[Gray and Lamport(2006)]%
        {10.1145/1132863.1132867}
\bibfield{author}{\bibinfo{person}{Jim Gray} {and} \bibinfo{person}{Leslie
  Lamport}.} \bibinfo{year}{2006}\natexlab{}.
\newblock \showarticletitle{Consensus on Transaction Commit}.
\newblock \bibinfo{journal}{\emph{ACM Trans. Database Syst.}}
  \bibinfo{volume}{31}, \bibinfo{number}{1} (\bibinfo{date}{March}
  \bibinfo{year}{2006}), \bibinfo{pages}{133–160}.
\newblock
\showISSN{0362-5915}
\urldef\tempurl%
\url{https://doi.org/10.1145/1132863.1132867}
\showDOI{\tempurl}


\bibitem[Grove et~al\mbox{.}(2019)]%
        {Res-X10:TOPLAS}
\bibfield{author}{\bibinfo{person}{David Grove}, \bibinfo{person}{Sara~S.
  Hamouda}, \bibinfo{person}{Benjamin Herta}, \bibinfo{person}{Arun Iyengar},
  \bibinfo{person}{Kiyokuni Kawachiya}, \bibinfo{person}{Josh Milthorpe},
  \bibinfo{person}{Vijay Saraswat}, \bibinfo{person}{Avraham Shinnar},
  \bibinfo{person}{Mikio Takeuchi}, {and} \bibinfo{person}{Olivier Tardieu}.}
  \bibinfo{year}{2019}\natexlab{}.
\newblock \showarticletitle{Failure Recovery in Resilient {X10}}.
\newblock \bibinfo{journal}{\emph{ACM Trans. Program. Lang. Syst.}}
  \bibinfo{volume}{41}, \bibinfo{number}{3}, Article \bibinfo{articleno}{15}
  (\bibinfo{date}{July} \bibinfo{year}{2019}), \bibinfo{numpages}{30}~pages.
\newblock
\showISSN{0164-0925}
\urldef\tempurl%
\url{https://doi.org/10.1145/3332372}
\showDOI{\tempurl}


\bibitem[K3D(2021)]%
        {k3d-website}
K3D \bibinfo{year}{2021}\natexlab{}.
\newblock \bibinfo{title}{K3D Project Website}.
\newblock \bibinfo{howpublished}{\url{https://k3d.io/}}.
\newblock


\bibitem[Kafka(2022)]%
        {kafka}
Kafka \bibinfo{year}{2022}\natexlab{}.
\newblock \bibinfo{title}{Apache Kafka Project Website}.
\newblock \bibinfo{howpublished}{\url{https://kafka.apache.org}}.
\newblock


\bibitem[KAR(2022)]%
        {kar-github-anon}
KAR \bibinfo{year}{2022}\natexlab{}.
\newblock \bibinfo{title}{KAR GitHub}.
\newblock \bibinfo{howpublished}{https://github.com/ibm/kar}.
\newblock


\bibitem[Microsoft(2016)]%
        {microsoft-azure-functions}
\bibfield{author}{\bibinfo{person}{Microsoft}.}
  \bibinfo{year}{2016}\natexlab{}.
\newblock \bibinfo{title}{Azure Functions}.
\newblock
\newblock
\urldef\tempurl%
\url{https://functions.azure.com/}
\showURL{%
\tempurl}


\bibitem[Microsoft(2018)]%
        {durable-functions}
\bibfield{author}{\bibinfo{person}{Microsoft}.}
  \bibinfo{year}{2018}\natexlab{}.
\newblock \bibinfo{title}{Durable Functions Website}.
\newblock
  \bibinfo{howpublished}{\url{https://docs.microsoft.com/en-us/azure/azure-functions/durable/}}.
\newblock


\bibitem[Moritz et~al\mbox{.}(2018)]%
        {222605}
\bibfield{author}{\bibinfo{person}{Philipp Moritz}, \bibinfo{person}{Robert
  Nishihara}, \bibinfo{person}{Stephanie Wang}, \bibinfo{person}{Alexey
  Tumanov}, \bibinfo{person}{Richard Liaw}, \bibinfo{person}{Eric Liang},
  \bibinfo{person}{Melih Elibol}, \bibinfo{person}{Zongheng Yang},
  \bibinfo{person}{William Paul}, \bibinfo{person}{Michael~I. Jordan}, {and}
  \bibinfo{person}{Ion Stoica}.} \bibinfo{year}{2018}\natexlab{}.
\newblock \showarticletitle{Ray: A Distributed Framework for Emerging {AI}
  Applications}. In \bibinfo{booktitle}{\emph{13th USENIX Symposium on
  Operating Systems Design and Implementation (OSDI 18)}}.
  \bibinfo{publisher}{USENIX Association}, \bibinfo{address}{Carlsbad, CA},
  \bibinfo{pages}{561--577}.
\newblock
\showISBNx{978-1-939133-08-3}
\urldef\tempurl%
\url{https://www.usenix.org/conference/osdi18/presentation/moritz}
\showURL{%
\tempurl}


\bibitem[Mukherjee et~al\mbox{.}(2019)]%
        {RSM:ecoop2019}
\bibfield{author}{\bibinfo{person}{Suvam Mukherjee},
  \bibinfo{person}{Nitin~John Raj}, \bibinfo{person}{Krishnan Govindraj},
  \bibinfo{person}{Pantazis Deligiannis}, \bibinfo{person}{Chandramouleswaran
  Ravichandran}, \bibinfo{person}{Akash Lal}, \bibinfo{person}{Aseem Rastogi},
  {and} \bibinfo{person}{Raja Krishnaswamy}.} \bibinfo{year}{2019}\natexlab{}.
\newblock \showarticletitle{{Reliable State Machines: A Framework for
  Programming Reliable Cloud Services}}. In \bibinfo{booktitle}{\emph{33rd
  European Conference on Object-Oriented Programming (ECOOP 2019)}}
  \emph{(\bibinfo{series}{Leibniz International Proceedings in Informatics
  (LIPIcs)}, Vol.~\bibinfo{volume}{134})},
  \bibfield{editor}{\bibinfo{person}{Alastair~F. Donaldson}} (Ed.).
  \bibinfo{publisher}{Schloss Dagstuhl--Leibniz-Zentrum fuer Informatik},
  \bibinfo{address}{Dagstuhl, Germany}, \bibinfo{pages}{18:1--18:29}.
\newblock
\showISBNx{978-3-95977-111-5}
\showISSN{1868-8969}
\urldef\tempurl%
\url{https://doi.org/10.4230/LIPIcs.ECOOP.2019.18}
\showDOI{\tempurl}


\bibitem[Orleans-5456(2019)]%
        {orleans-call-chain-bug}
Orleans-5456 \bibinfo{year}{2019}\natexlab{}.
\newblock \bibinfo{title}{Simple call chain reentrancy deadlock A->B->C->A}.
\newblock
  \bibinfo{howpublished}{\url{https://github.com/dotnet/orleans/issues/5456}}.
\newblock


\bibitem[Orleans-7397(2021)]%
        {orleans-call-chain-removed}
Orleans-7397 \bibinfo{year}{2021}\natexlab{}.
\newblock \bibinfo{title}{Description of call chain reentrancy is missing}.
\newblock
  \bibinfo{howpublished}{\url{https://github.com/dotnet/orleans/issues/7397}}.
\newblock


\bibitem[Redis(2022)]%
        {redis}
Redis \bibinfo{year}{2022}\natexlab{}.
\newblock \bibinfo{title}{Redis Project Website}.
\newblock \bibinfo{howpublished}{\url{https://redis.io}}.
\newblock


\bibitem[Sang et~al\mbox{.}(2020)]%
        {Sang:OOPSLA2020}
\bibfield{author}{\bibinfo{person}{Bo Sang}, \bibinfo{person}{Patrick Eugster},
  \bibinfo{person}{Gustavo Petri}, \bibinfo{person}{Srivatsan Ravi}, {and}
  \bibinfo{person}{Pierre-Louis Roman}.} \bibinfo{year}{2020}\natexlab{}.
\newblock \showarticletitle{Scalable and Serializable Networked Multi-Actor
  Programming}.
\newblock \bibinfo{journal}{\emph{Proc. ACM Program. Lang.}}
  \bibinfo{volume}{4}, \bibinfo{number}{OOPSLA}, Article
  \bibinfo{articleno}{198} (\bibinfo{date}{nov} \bibinfo{year}{2020}),
  \bibinfo{numpages}{30}~pages.
\newblock
\urldef\tempurl%
\url{https://doi.org/10.1145/3428266}
\showDOI{\tempurl}


\bibitem[Sato et~al\mbox{.}(2012)]%
        {Sato12sc}
\bibfield{author}{\bibinfo{person}{Kento Sato}, \bibinfo{person}{Naoya
  Maruyama}, \bibinfo{person}{Kathryn Mohror}, \bibinfo{person}{Adam Moody},
  \bibinfo{person}{Todd Gamblin}, \bibinfo{person}{Bronis~R. de Supinski},
  {and} \bibinfo{person}{Satoshi Matsuoka}.} \bibinfo{year}{2012}\natexlab{}.
\newblock \showarticletitle{{Design and modeling of a non-blocking
  checkpointing system}}. In \bibinfo{booktitle}{\emph{Proc. International
  Conference for High Performance Computing, Networking, Storage and Analysis
  2012 (SC~'12)}}.
\newblock


\bibitem[Setty et~al\mbox{.}(2016)]%
        {199350}
\bibfield{author}{\bibinfo{person}{Srinath Setty}, \bibinfo{person}{Chunzhi
  Su}, \bibinfo{person}{Jacob~R. Lorch}, \bibinfo{person}{Lidong Zhou},
  \bibinfo{person}{Hao Chen}, \bibinfo{person}{Parveen Patel}, {and}
  \bibinfo{person}{Jinglei Ren}.} \bibinfo{year}{2016}\natexlab{}.
\newblock \showarticletitle{Realizing the Fault-Tolerance Promise of Cloud
  Storage Using Locks with Intent}. In \bibinfo{booktitle}{\emph{12th {USENIX}
  Symposium on Operating Systems Design and Implementation ({OSDI} 16)}}.
  \bibinfo{publisher}{{USENIX} Association}, \bibinfo{address}{Savannah, GA},
  \bibinfo{pages}{501--516}.
\newblock
\showISBNx{978-1-931971-33-1}
\urldef\tempurl%
\url{https://www.usenix.org/conference/osdi16/technical-sessions/presentation/setty}
\showURL{%
\tempurl}


\bibitem[Sharma et~al\mbox{.}(2016)]%
        {10.1016/j.jnca.2016.08.010}
\bibfield{author}{\bibinfo{person}{Yogesh Sharma}, \bibinfo{person}{Bahman
  Javadi}, \bibinfo{person}{Weisheng Si}, {and} \bibinfo{person}{Daniel Sun}.}
  \bibinfo{year}{2016}\natexlab{}.
\newblock \showarticletitle{Reliability and Energy Efficiency in Cloud
  Computing Systems}.
\newblock \bibinfo{journal}{\emph{J. Netw. Comput. Appl.}}
  \bibinfo{volume}{74}, \bibinfo{number}{C} (\bibinfo{date}{Oct.}
  \bibinfo{year}{2016}), \bibinfo{pages}{66–85}.
\newblock
\showISSN{1084-8045}
\urldef\tempurl%
\url{https://doi.org/10.1016/j.jnca.2016.08.010}
\showDOI{\tempurl}


\bibitem[Shipping(2018a)]%
        {icg-shipping-arch}
Shipping \bibinfo{year}{2018}\natexlab{a}.
\newblock \bibinfo{title}{Shipping Container Application Architecture}.
\newblock \bibinfo{howpublished}{URL omitted for double-blind reviewing}.
\newblock


\bibitem[Shipping(2018b)]%
        {icg-shipping-impl}
Shipping \bibinfo{year}{2018}\natexlab{b}.
\newblock \bibinfo{title}{Shipping Container Application Implementation}.
\newblock \bibinfo{howpublished}{URL omitted for double-blind reviewing}.
\newblock


\bibitem[Vishwanath and Nagappan(2010)]%
        {10.1145/1807128.1807161}
\bibfield{author}{\bibinfo{person}{Kashi~Venkatesh Vishwanath} {and}
  \bibinfo{person}{Nachiappan Nagappan}.} \bibinfo{year}{2010}\natexlab{}.
\newblock \showarticletitle{Characterizing Cloud Computing Hardware
  Reliability}. In \bibinfo{booktitle}{\emph{Proceedings of the 1st ACM
  Symposium on Cloud Computing}} (Indianapolis, Indiana, USA)
  \emph{(\bibinfo{series}{SoCC '10})}. \bibinfo{publisher}{Association for
  Computing Machinery}, \bibinfo{address}{New York, NY, USA},
  \bibinfo{pages}{193–204}.
\newblock
\showISBNx{9781450300360}
\urldef\tempurl%
\url{https://doi.org/10.1145/1807128.1807161}
\showDOI{\tempurl}


\bibitem[White(2009)]%
        {White:2009:HDG:1717298}
\bibfield{author}{\bibinfo{person}{Tom White}.}
  \bibinfo{year}{2009}\natexlab{}.
\newblock \bibinfo{booktitle}{\emph{Hadoop: The Definitive Guide}
  (\bibinfo{edition}{1st} ed.)}.
\newblock \bibinfo{publisher}{O'Reilly Media, Inc.}
\newblock
\showISBNx{0596521979, 9780596521974}


\bibitem[Zaharia et~al\mbox{.}(2012)]%
        {Zaharia2012:USENIX}
\bibfield{author}{\bibinfo{person}{Matei Zaharia}, \bibinfo{person}{Mosharaf
  Chowdhury}, \bibinfo{person}{Tathagata Das}, \bibinfo{person}{Ankur Dave},
  \bibinfo{person}{Justin Ma}, \bibinfo{person}{Murphy McCauley},
  \bibinfo{person}{Michael~J Franklin}, \bibinfo{person}{Scott Shenker}, {and}
  \bibinfo{person}{Ion Stoica}.} \bibinfo{year}{2012}\natexlab{}.
\newblock \showarticletitle{Resilient distributed datasets: A fault-tolerant
  abstraction for in-memory cluster computing}. In
  \bibinfo{booktitle}{\emph{Proc. 9th USENIX Symposium on Networked Systems
  Design and Implementation (NSDI 12)}}. USENIX Association,
  \bibinfo{pages}{15--28}.
\newblock


\bibitem[Zhang et~al\mbox{.}(2020)]%
        {258880}
\bibfield{author}{\bibinfo{person}{Haoran Zhang}, \bibinfo{person}{Adney
  Cardoza}, \bibinfo{person}{Peter~Baile Chen}, \bibinfo{person}{Sebastian
  Angel}, {and} \bibinfo{person}{Vincent Liu}.}
  \bibinfo{year}{2020}\natexlab{}.
\newblock \showarticletitle{Fault-tolerant and transactional stateful
  serverless workflows}. In \bibinfo{booktitle}{\emph{14th {USENIX} Symposium
  on Operating Systems Design and Implementation ({OSDI} 20)}}.
  \bibinfo{publisher}{{USENIX} Association}, \bibinfo{pages}{1187--1204}.
\newblock
\showISBNx{978-1-939133-19-9}
\urldef\tempurl%
\url{https://www.usenix.org/conference/osdi20/presentation/zhang-haoran}
\showURL{%
\tempurl}


\end{thebibliography}

\end{document}